\RequirePackage{rotfloat}
\documentclass[fleqn,usenatbib]{mnras}

\usepackage{newtxtext,newtxmath}

\usepackage[T1]{fontenc}

\DeclareRobustCommand{\VAN}[3]{#2}
\let\VANthebibliography\thebibliography
\def\thebibliography{\DeclareRobustCommand{\VAN}[3]{##3}\VANthebibliography}

\usepackage{commath, graphicx, bbold, endnotes, graphicx, subfigure, multirow, afterpage}

\usepackage{graphicx}
\usepackage{caption}
\usepackage{setspace}

\newcommand\data{\boldsymbol{v}}

\newcommand\gains{\boldsymbol{g}}

\newcommand\electricfield{\boldsymbol{E}}

\newcommand\racoord{R}
\newcommand\deccoord{D}

\newcommand\rotangle{\beta}
\newcommand\rotmeas{R}
\newcommand\wavelength{\lambda}
\newcommand\skypos{\boldsymbol{\theta}}

\newcommand\obsindex{n}
\newcommand\radweight{W}

\title[A Map of Diffuse Radio Emission at 182 MHz]{A Map of Diffuse Radio Emission at 182 MHz to Enhance Epoch of Reionization Observations in the Southern Hemisphere}

\author[R. Byrne et al.]{
Ruby Byrne,$^{1,2}$\thanks{E-mail: rbyrne@caltech.edu} 
Miguel F.~Morales,$^{2}$ 
Bryna Hazelton,$^{2,3}$ 
Ian Sullivan,$^{4}$ 
Nichole Barry,$^{5,6}$
\newauthor
Christene Lynch,$^{6,7}$  
Jack L.~B.~Line,$^{6,7}$
and Daniel C.~Jacobs$^{8}$
\\
$^{1}$Astronomy Department, California Institute of Technology, 1200 E California Blvd, Pasadena, CA, 91125, USA\\
$^{2}$Physics Department, University of Washington, 3910 15th Ave NE, Seattle, WA, 98195-1560, USA\\
$^{3}$eScience Institute, University of Washington, 3910 15th Ave NE, Seattle, WA, 98195-1560, USA\\
$^{4}$Astronomy Department, University of Washington, 3910 15th Ave NE, Seattle, WA, 98195-1560, USA\\
$^{5}$School of Physics, University of Melbourne, Parkville, VIC 3010, Australia\\
$^{6}$Australian Research Council Centre of Excellence for All Sky Astrophysics in 3 Dimensions (ASTRO 3D), Australia\\
$^{7}$International Centre for Radio Astronomy Research, Curtin University, 1 Turner Avenue, Bentley WA 6102, Australia\\
$^{8}$School of Earth and Space Exploration, Arizona State University, PO Box 876004, Tempe, AZ, 85287-6004, USA
}

\date{Accepted XXX. Received YYY; in original form ZZZ}

\begin{document}
\label{firstpage}
\pagerange{\pageref{firstpage}--\pageref{lastpage}}
\maketitle

\begin{abstract}
We present a broadband map of polarized diffuse emission at 167-198 MHz developed from data from the Murchison Widefield Array (MWA). The map is designed to improve visibility simulation and precision calibration for 21 cm Epoch of Reionization (EoR) experiments. It covers a large swath---11,000 sq.\ deg.---of the Southern Hemisphere sky in all four Stokes parameters and captures emission on angular scales of $1^\circ$ to $9^\circ$. The band-averaged diffuse structure is predominantly unpolarized but has significant linearly polarized structure near RA = 0 h. We evaluate the accuracy of the map by combining it with the GLEAM catalog and simulating an observation from the MWA, demonstrating that the accuracy of the short baselines (6.1-50 wavelengths) now approaches the accuracy of the longer baselines typically used for EoR calibration. We discuss how to use the map for visibility simulation for a variety of interferometric arrays. The map has potential to improve calibration accuracy for experiments such as the Hydrogen Epoch of Reionization Array (HERA) and the forthcoming Square Kilometre Array (SKA) as well as the MWA.
\end{abstract}




\section{Introduction}

The low-frequency radio sky is bright at large scales, yet our understanding of this diffuse emission is limited. Mapping this emission can illuminate the structure of the Milky Way's interstellar medium (ISM), and measurement of its polarization modes probes galactic magnetic fields \citep{McKee2007, Jelic2010, Bernardi2013, Lenc2016}. Furthermore, diffuse mapping can aid in the precision calibration of radio cosmology instruments, enabling deeper limits on the highly redshifted 21 cm signal.

Motivated by the extreme calibration precision required for low-frequency radio cosmology \citep{Barry2016, Trott2016a, Patil2016, Byrne2019}, we present a high-fidelity map of broadband diffuse emission at 182 MHz across much of the Southern Hemisphere sky. The map covers 11,000 sq.\ deg., including approximate Right Ascensions (RAs) of $-3.5$ to $9.5$ h and Declinations (Decs) of $-62^\circ$ to $10^\circ$. It represents angular scales of approximately $1^\circ$ to $9^\circ$ in all four Stokes parameters. The map is measured across a 31 MHz frequency continuum and captures Stokes Q and U emission with low ($< 2$ rad/m$^2$) rotation measure (RM) magnitudes. Compact sources have been removed from the data, meaning that the map can be combined with existing compact source catalogs to produce an expanded and improved sky model for calibration.

The map is produced from data from the Murchison Widefield Array (MWA), a radio interferometer in the Western Australian outback, in its Phase I configuration \citep{Tingay2013}. This instrument is particularly well-suited to diffuse mapping because of it high \textit{uv} coverage at short baselines and good resolution from long baselines. The resolution allows for localizing and removing compact sources, and the density of short baselines allows for high sensitivity measurements of diffuse emission. Furthermore, across a large frequency continuum the short baselines sample nearly every \textit{uv} mode shorter than 50 wavelengths. This enables us to produce maps of diffuse emission in physical units of surface brightness without deconvolving. The maps can be transferred to other instruments for visibility simulation.

Existing low-frequency diffuse maps include \citealt{Haslam1981, Haslam1982, Bennett2003, DeOliveira-Costa2008, Remazeilles2015, Zheng2017, Eastwood2018}; and \citealt{Wolleben2019}. These maps have minimal, if any, compact source subtraction, meaning that they cannot be combined with deep catalogs of compact sources. All but \citealt{Wolleben2019} are unpolarized, omitting the bright linearly polarized emission seen on large angular scales. As a result, these maps do not have the precision required to enable highly accurate cosmological calibration. In contrast, we present a map with deep compact source removal and accurate polarization reconstruction in the widefield limit.

We quantify the map's accuracy by measuring the discrepancy between visibilities simulated from the map and true data from an MWA observation. We find that the diffuse map reconstructs a the majority of the observation's visibility power on short baselines. The map improves short baseline modeling to accuracy levels nearly equivalent to those of long baseline modeling with state of the art compact source sky models.

We expect that the map presented here will improve high-precision calibration for 21 cm cosmology experiments with not only the MWA itself but also other Southern Hemisphere interferometers such as the Hydrogen Epoch of Reionization Array (HERA) in South Africa \citep{DeBoer2017} and future instruments such as the Square Kilometre Array (SKA). HERA is highly compact, dominated by short baseline measurements, and is therefore particularly sensitive to diffuse structure in calibration. Improved diffuse modeling could also enable new short-baseline calibration approaches that mitigate systematic errors \citep{Ewall-Wice2017}.

In this paper \S\ref{s:methods} describes the data processing procedure and \S\ref{s:results} presents the resulting polarized diffuse map. \S\ref{s:angular_scales} describes the angular scales over which we have confidence in the results, \S\ref{s:validation} quantifies the effect of the map on improving visibility simulation for an MWA observation, and \S\ref{s:discussion} presents a discussion of how to use this diffuse map for visibility simulation.

\section{Methods}
\label{s:methods}

In this section we describe the analysis process and techniques used to achieve a high-fidelity polarized map of diffuse emission. The bulk of the analysis is performed with the Fast Holographic Deconvolution (\textsc{fhd}) software package\footnote{\texttt{https://github.com/EoRImaging/FHD}} \citep{Sullivan2012, Barry2019a}. We analyze all data in the Amazon Web Services (AWS) cloud using the cloud computing pipeline described in \citealt{Byrne2021a}.

\subsection{The Data}

\begin{figure*}
    \centering
    \subfigure[]{
    \includegraphics[width=1.5\columnwidth]{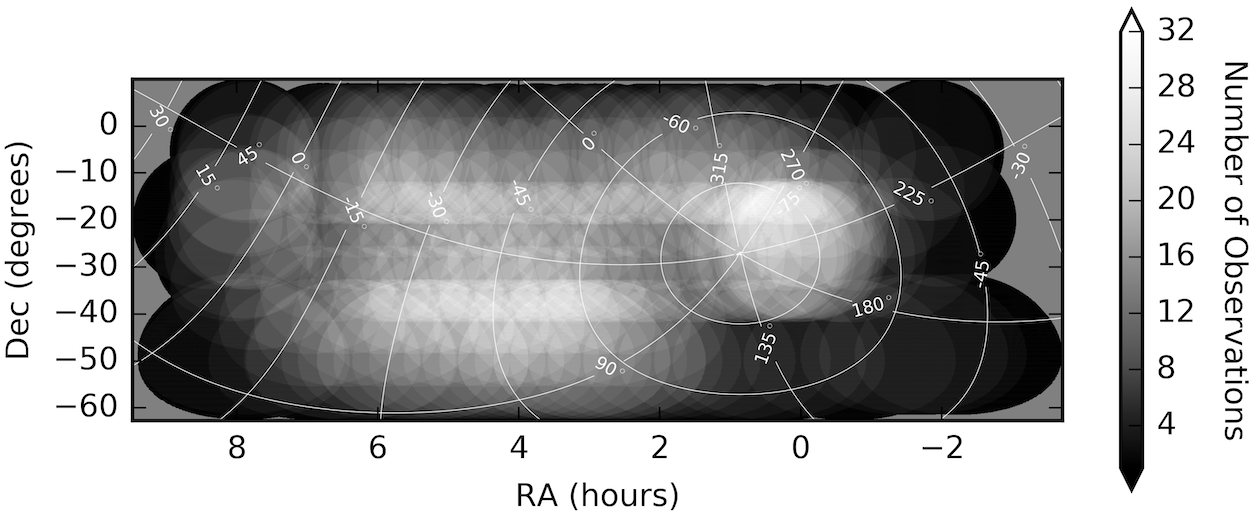}
    \label{fig:nsamples}}
    \subfigure[]{
    \includegraphics[width=1.5\columnwidth]{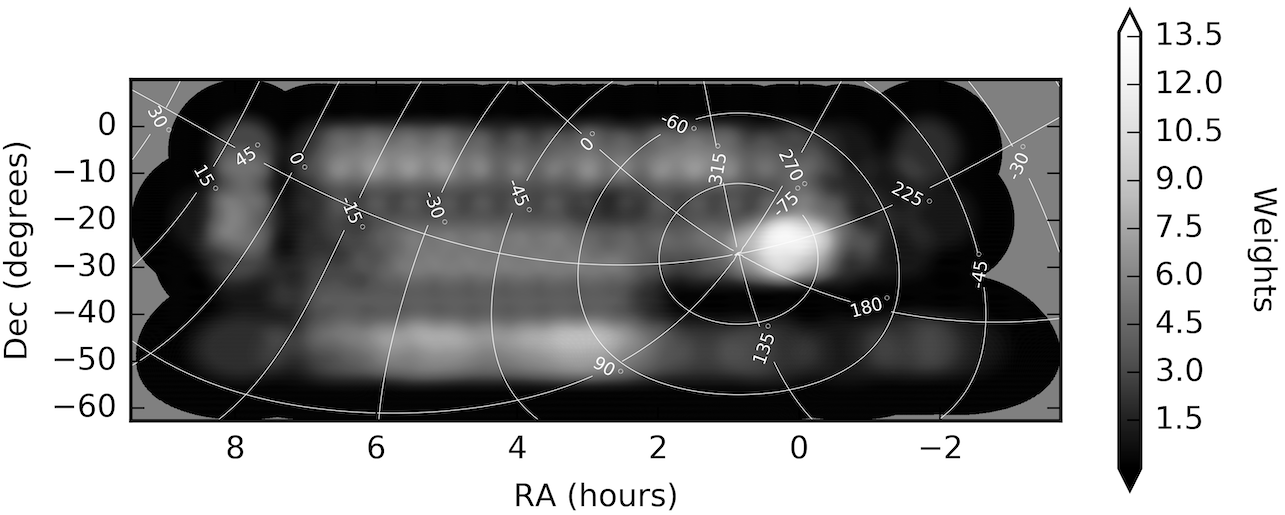}
    \label{fig:weighting}
    }
    \caption{Plots of the observational coverage of the map. Observations are chosen to overlap each field on the sky. (a) depicts the number of observations that contribute to each field. The observations are averaged with a weighting function given by Equation \ref{eq:weighting_function}. (b) depicts the total weight in each field on the sky, given this weighting scheme. The galactic coordinates are indicated in white. We use an especially large sample of observations of the ``EoR-0'' field centred at RA 0 h, Dec.\ $-27^\circ$ because it is of particular interest to the MWA's EoR team and displays strong linearly polarized signal. Just south of that field, at a declination of about $-50^\circ$, the map has a dearth of observational coverage owing to scheduling limitations during the data acquisition.}
\label{fig:obs_coverage}
\end{figure*}

We present a map comprising 5.8 hours of data from the MWA Phase I, taken November 5-11, 2015. The data consists of 172 individual observations that are approximately 2 minutes (112 seconds) in length. 
This time interval is sufficiently short to enable each observation to be processed as a fringe stopped snapshot image.

The observational fields are chosen to overlap across the entire survey field, plotted in Figure \ref{fig:obs_coverage}. Each field is observed with a variety of elevation angles to maximize \textit{uv} rotation and reduce sidelobe confusion.

The observations cover frequencies of 167-198 MHz and are processed as continuum images averaged across the full 31 MHz frequency range. This further fills the \textit{uv} plane, reducing sidelobe confusion and providing an accurate measurement of diffuse power on all \textit{uv} modes. However, it also means that polarized emission with substantial frequency evolution will depolarize through frequency averaging. We therefore expect to only measure polarized emission with low rotation measure magnitudes.

Data pre-processing is performed with the \textsc{cotter} package\footnote{\texttt{https://github.com/MWATelescope/cotter}} \citep{Offringa2015}. This reduces the data volume by frequency-averaging to 80 kHz and time-averaging to 2 seconds. We perform Radio-Frequency Interference (RFI) excision in two steps. Initial RFI flagging is performed with \textsc{aoflagger}\footnote{\texttt{https://sourceforge.net/projects/aoflagger/}}, which is contained within the \textsc{cotter} pre-processing package. We subsequently perform additional RFI flagging with \textsc{ssins}\footnote{\texttt{https://github.com/mwilensky768/SSINS}} \citep{Wilensky2019}.

\subsection{Calibration}
\label{s:calibration}

We calibrate with \textsc{fhd}'s sky-based calibration pipeline, leveraging the precision calibration techniques developed for EoR science \citep{Barry2016, Beardsley2016, Barry2019a, Barry2019b, Li2019}.

The calibration sky model is based on the GLEAM catalog \citep{Hurley-Walker2017}. As GLEAM omits some of the brightest sources in the field, we supplement the catalog with additional source models. We use point source models for 3C161, 3C409, and Cassiopeia A and multi-component extended source models for Centaurus A, Hera A, Hydra A, Pictor A, and Virgo A (White, private communication). We additionally include an extended source model for Fornax A produced with \textsc{fhd} (Carroll, private communication). All extended source models use point-like sub-components to represent the sources' extended structure.

We use \textsc{fhd} to simulate visibilities for calibration from this supplemented GLEAM catalog. We calibrate with all baselines longer than 50 wavelengths. Shorter baselines are poorly modeled by the compact source catalog, and we omit them from calibration to avoid bias in our map estimation.

\textsc{fhd}'s calibration calculates a complex gain per antenna polarization. We denote the two antenna polarization modes $p$ and $q$, such that calibration calculates gains $\gains_p$ and $\gains_q$ for each antenna, frequency interval, and time step. The MWA has well-isolated signal paths and experiences minimal cross-talk between the instrumental polarizations. We therefore do not fit calibration parameters that mix $p$ and $q$ \citep{Hamaker1996a, Sault1996}.

Previous unpolarized analyses with \textsc{fhd} calibrate using only the single-polarization visibilities, i.e.\ $\data_{pp}$ and $\data_{qq}$ \citep{Barry2019a}. This does not enable fully polarized imaging because it leaves the phase relationship between $\gains_p$ and $\gains_q$ unconstrained. We therefore include an additional calibration step in which we constrain this degeneracy with the cross-polarization visibilities $\data_{pq}$ and $\data_{qp}$ (Byrne thesis, 2021). We fit all other calibration degrees-of-freedom with only the single-polarization visibilities. For further discussion of polarized calibration, see \citealt{Sault1996, Lenc2017, Gehlot2018, Dillon2018, Byrne2021b}.

Calibration constrains the observations' overall flux scale. The resulting calibrated images are therefore normalized to the flux scale of the GLEAM catalog, which is in turn referenced to the Very Large Array Low-frequency Sky
Survey Redux (VLSSr; \citealt{Lane2014}), the Molonglo Reference Catalogue (MRC; \citealt{Large1981}), the National Radio Astronomy Observatory Very Large Array Sky Survey (NVSS; \citealt{Condon1998}), and \citealt{Baars1977} \citep{Hurley-Walker2017}.

\subsection{Imaging and Compact Source Removal}
\label{s:imaging}

\textsc{fhd} performs high-fidelity widefield imaging based on the optimal mapmaking formalism \citep{Bhatnagar2008, Morales2009}. It uses the instrumental response kernel to grid visibilities to a \textit{uv} plane, which we pixelate at half-wavelength spacing. This naturally yields an accurate horizon-to-horizon image reconstruction without requiring faceting across the instrumental field-of-view, enabling measurement of large-scale structure spanning several angular degrees.

To enable this work, we updated \textsc{fhd} to perform fully-polarized imaging (Byrne thesis, 2021). Images are reconstructed in the non-orthogonal ``instrumental'' polarization basis, aligned with the polarization vectors of maximal instrumental response. We then use the instrument's polarized response model, or Jones matrix \citep{Jones1941}, to translate the instrumental polarization modes to Stokes polarization modes, following the formalism developed in \citealt{Hamaker1996a, Sault1996, Hamaker1996b, Hamaker2000, Hamaker2006, Ord2010}. We define the Stokes parameters $I$, $Q$, $U$, and $V$ as
\begin{equation}
    \begin{bmatrix}
    I(\skypos) \\
    Q(\skypos) \\
    U(\skypos) \\
    V(\skypos)
    \end{bmatrix} = \begin{bmatrix}
	1 & 1 & 0 & 0 \\
	1 & -1 & 0 & 0 \\
	0 & 0 & 1 & 1 \\
	0 & 0 & i & -i \\
	\end{bmatrix} \begin{bmatrix}
    \langle |\electricfield_\racoord(\skypos)|^2 \rangle \\ \langle |\electricfield_\deccoord(\skypos)|^2 \rangle \\
    \langle \electricfield_\racoord(\skypos) \electricfield_\deccoord^*(\skypos) \rangle \\
    \langle \electricfield_\racoord^*(\skypos) \electricfield_\deccoord(\skypos) \rangle
    \end{bmatrix}.
\label{eq:stokes_def}
\end{equation}
Here $\electricfield_\racoord$ and $\electricfield_\deccoord$ represent the RA- and Dec.-aligned electric field vectors, respectively. $\skypos$ denotes a position on the sky, and the angle brackets $\langle \rangle$ indicate the time average.

We crop each image at a diameter of $30^\circ$ to limit the images to regions of high sensitivity. Next, we perform a data quality cut based on calibration success. We omit observations with clear calibration systematics, usually apparent as high Stokes V power across the image, from further analysis. The effect shows some time correlation but is not field-dependent: observations of the same field experience variable calibration success. We expect that poor calibration solutions could result from unmitigated RFI contamination or ionospheric activity \citep{Jordan2017}.

Next, we remove compact sources with \textsc{fhd}'s deconvolution algorithm \citep{Sullivan2012}. We expect compact sources to have negligible intrinsic polarization \citep{Ord2010, Bernardi2013}, and we therefore restrict our source models to Stokes I only in this step. Compared to simply modeling and subtracting the calibration catalog, deconvolving compact sources with \textsc{fhd} produces better removal of extended sources and mitigation of source intensity errors from beam response mismodeling. We remove about 10,000 sources from each observation.

\subsection{Correcting for Polarization Leakage from Beam Modeling Errors}
\label{s:pol_leakage_correction}

Errors in the polarized instrumental response model lead to polarization mode-mixing \citep{Ord2010, Sutinjo2015}. Our observations are dominated by power from unpolarized point-like sources \citep{Bernardi2013}, and we see evidence of beam modeling errors in the form of significant point source power reconstructed in Stokes Q and U. As \textsc{fhd}'s deconvolution algorithm removes sources in Stokes I only, this power persists through source removal and contaminates the Stokes Q and U data products. We do not see high levels of Stokes I to V leakage.

\begin{figure*}
\centering
\includegraphics[width=2\columnwidth]{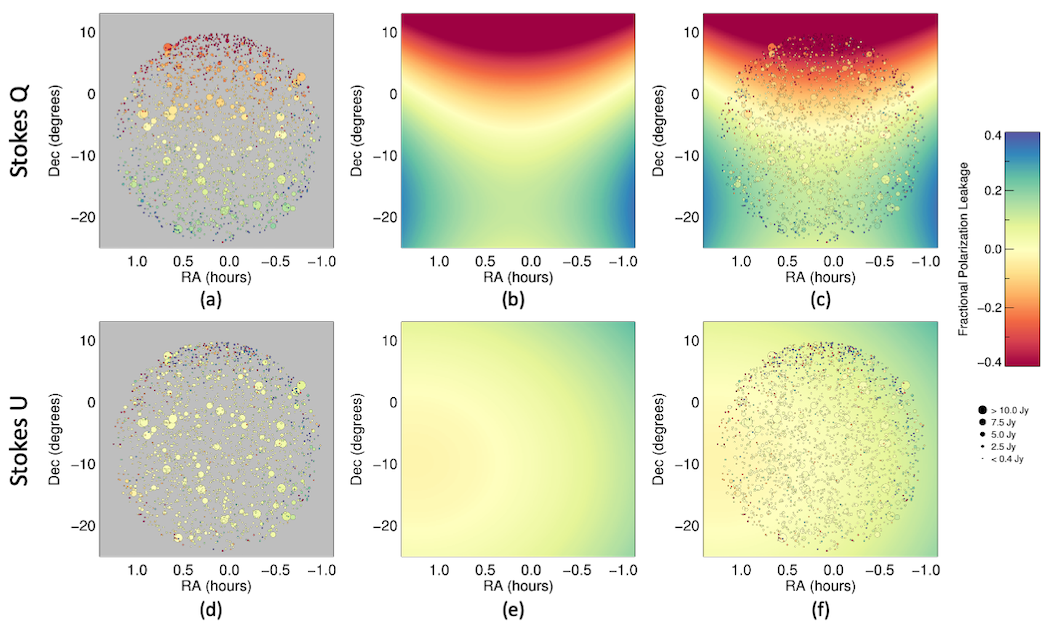}
\caption{Errors in the polarized instrumental response model lead to significant observed compact source power in Stokes Q (top row) and U (bottom row). In the left column we plot the observed Stokes Q and U polarization fraction of the 2,000 brightest sources removed from a single observation. The point size corresponds to the source intensity, and its colour corresponds to the apparent polarization fraction. In the centre column we display a second-order polynomial surface fit that we use to mitigate residual point source power in Stokes Q and U. In the right column we combine the sources and surface fit.}
\label{fig:polarization_correction}
\end{figure*}

To mitigate residual point source power in Stokes Q and U, we follow techniques similar to those presented in \citealt{Lenc2017} to fit and subtract polarization leakage. We apply the correction to compact sources only and do not adjust the measured diffuse emission based on this fit.

\begin{figure*}
\centering
\includegraphics[width=1.5\columnwidth]{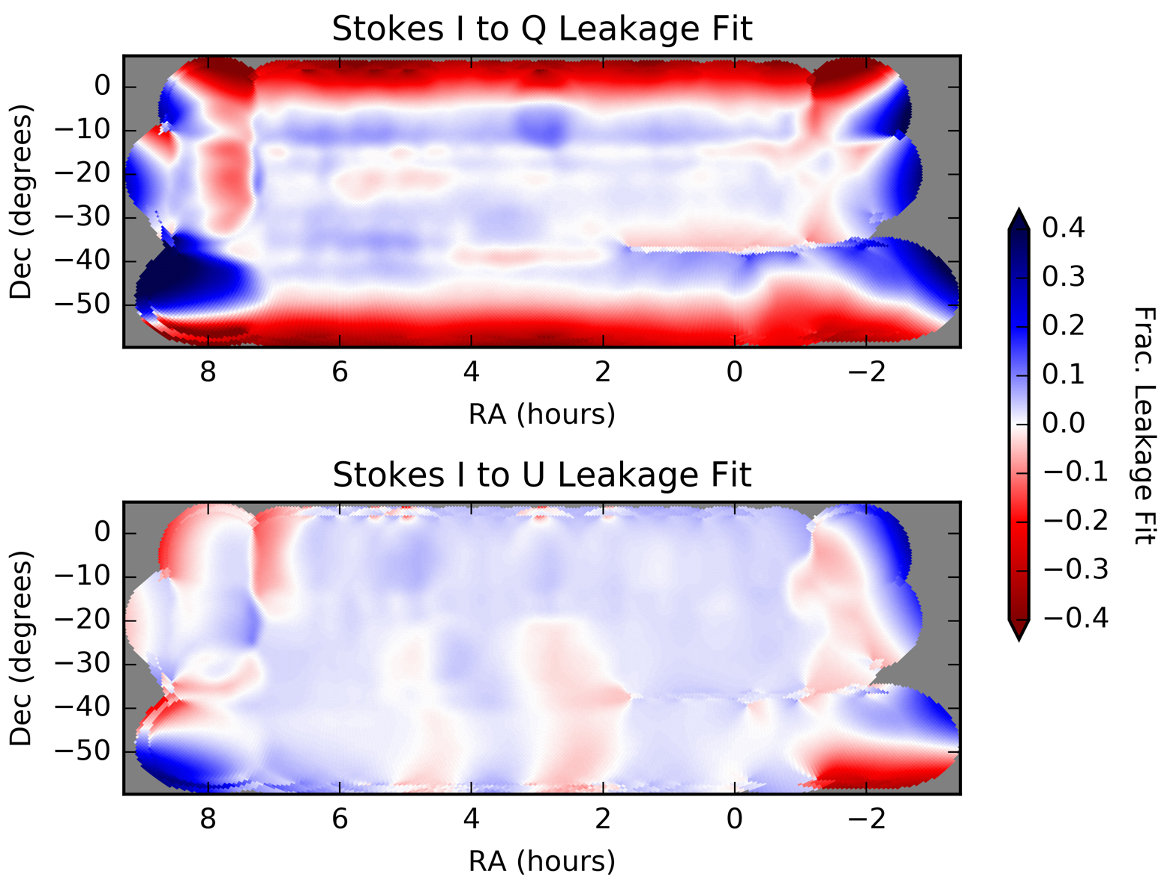}
\caption{Estimate of Stokes I to Q and U leakage across the full map. The estimate is derived from the fit leakage surfaces calculated for each 2-minute observation, as depicted in Figure \ref{fig:polarization_correction}. We combine the observations' fit surfaces across the full map using the weighting presented in Equation \ref{eq:weighting_function}. Leakage is high toward the edges of the map, particularly in Stokes Q, as these field have lower observational coverage (plotted in Figure \ref{fig:obs_coverage}) and are not observed with many unique pointings.}
\label{fig:frac_leakage_estimate}
\end{figure*}

For every observation, we calculate the Stokes Q and U polarization components at each of the locations of the 2,000 brightest sources removed. We then convert that measurement into an apparent Stokes Q and U polarization fraction. We use these values to fit a second-order polynomial surface across the observation field-of-view; the fit is weighted by source intensity, preferentially fitting to the brightest observed sources. Figure \ref{fig:polarization_correction} presents the result of this fit for an example observation. Figure \ref{fig:frac_leakage_estimate} estimates the leakage fraction across the full map.

We use the fit polynomial surface for each observation to modify the catalogs of sources removed with \textsc{fhd}'s deconvolution. Each source is modified to reflect its apparent polarization mode according to the surface fit, up to 40\% polarization fraction each in Stokes Q and U. We then model visiblities from those catalogs and subtract them from the data. This technique effectively removes residual point source power from Stokes Q and U, yielding cleaner polarized maps that better highlight diffuse emission.

At this point we could choose to use the same fit polynomial surface to modify the diffuse maps themselves. However, this approach would make the diffuse maps highly sensitive to small errors in the polarization leakage fit. The second-order polynomial surfaces represent structure on angular scales equivalent to those of the diffuse emission, and errors could significantly bias our diffuse measurement. Furthermore, as we capture Stokes I to Q and U leakage only, this modification would not capture other forms of polarization leakage.

\subsection{Normalization and Combining Observations}
\label{s:normalization_and_averaging}

\begin{figure}
\centering
\includegraphics[height=2.2in]{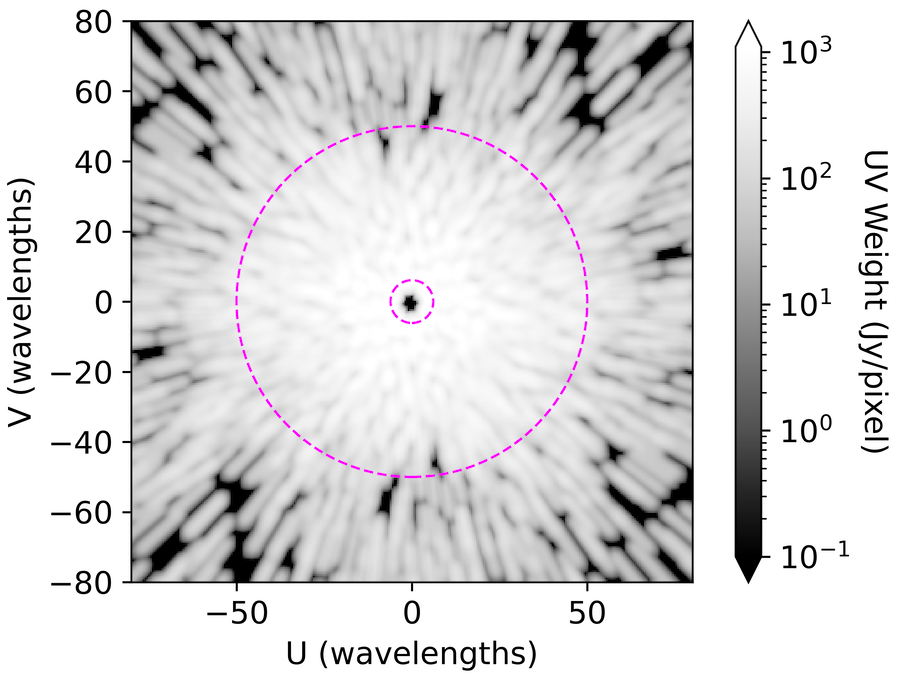}
\caption{Plot of the \textit{uv} measurement coverage for a zenith observation from the MWA Phase I for the 167-198 MHz frequency range. The large dashed magenta contour has a radius of 50 wavelengths, represented the longest baselines used in the diffuse map presented in this paper. Across this 31 MHz frequency continuum, the MWA samples nearly every \textit{uv} point below 50 wavelengths. The small dashed magenta contour has a radius of 6.1 wavelengths, representing the extent of our confidence in the diffuse map at large angular scales. In the optimal mapmaking formalism, this measurement coverage is known as the ``\textit{uv} weights'' (see \citealt{Barry2019a} for further discussion).}
\label{fig:uv_weights}
\end{figure}

After removing unpolarized point sources and mitigating point source polarization leakage in Stokes Q and U, we re-image each observation with short baselines only, omitting baselines longer than 50 wavelengths. This re-imaging accomplishes two things. First, it highlights diffuse emission by mapping large-scale structure only. Secondly, it limits the imaged data to regions of the \textit{uv} plane where the MWA samples virtually every \textit{uv} point (see Figure \ref{fig:uv_weights}). 

Imaging in this regime of near-complete \textit{uv} sampling is known as deconvolution-free imaging because it enables faithful reconstruction of the sky signal without deconvolving. This is a necessary condition for accurately imaging diffuse emission. Deconvolution inherently requires that the signal is sparse, with fewer components than independent measurements. For example, the deconvolution algorithm we use to remove compact sources, described above in \S\ref{s:imaging}, assumes that the signal is localized at discrete points on the sky. Diffuse emission is not sparse and produces an independent measurement at all \textit{uv} modes. A high-fidelity measurement of diffuse emission therefore requires a measurement of nearly all \textit{uv} modes.

For the MWA, achieving near-complete \textit{uv} sampling below 50 wavelengths requires imaging across a frequency continuum. The \textit{uv} plane is not well-sampled at any single frequency within our 31 MHz band. However, combining measurements across the full frequency range fills the \textit{uv} plane and offers improved \textit{uv} measurement coverage. 

By reconstructing diffuse emission in the deconvolution-free regime, we produce images with well-defined normalization on angular scales of $1^\circ$ to $9^\circ$ (see \S\ref{s:angular_scales}). This normalization is independent of the instrumental beam response (Byrne thesis, 2021), and the resulting maps can thereby be transferred to other instruments for visibility modeling. We present the diffuse map in surface brightness units of Janskys per steradian (Jy/sr).

Next, we average all observations to produce a single polarized map across the observed fields. The observations are averaged with a position-dependent tapered cosine (or Tukey) weighting $\radweight_\obsindex(\skypos)$ given by
\begin{equation}
    \radweight_\obsindex(\skypos) = \begin{cases}
    1, & \text{for $|\skypos - \skypos_\obsindex| < 4^\circ$} \\
    \cos \left( \frac{\pi}{12^\circ} |\skypos - \skypos_\obsindex| - \frac{\pi}{3} \right), & \text{for $4^\circ \le |\skypos - \skypos_\obsindex| \le 10^\circ$} \\
    0, & \text{for $|\skypos - \skypos_\obsindex| > 10^\circ$}
    \end{cases}.
\label{eq:weighting_function}
\end{equation}
Here $\obsindex$ indexes the observation and $\skypos$ represents the position on the sky. $\skypos_\obsindex$ denotes the position of the beam centre of observation $\obsindex$. This weighting reflects our greater confidence near the beam centre of each observation. Figure \ref{fig:weighting} depicts the total weight of each field across the map.

\subsection{Correcting for Ionospheric Faraday Rotation}
\label{s:ionosphere}

Propagation through intervening magnetized plasma induces Faraday rotation in linearly polarized emission. In particular, the ionosphere induces time- and direction-dependent Faraday rotation. This rotation mixes the Stokes Q and U polarization modes of measured radiation and causes depolarization across the measured frequency continuum and observations in the combined map \citep{Ord2010}. To mitigate these effects we correct for ionospheric Faraday rotation.

\begin{figure}
\centering
\subfigure[]{
\includegraphics[width=3.3in]{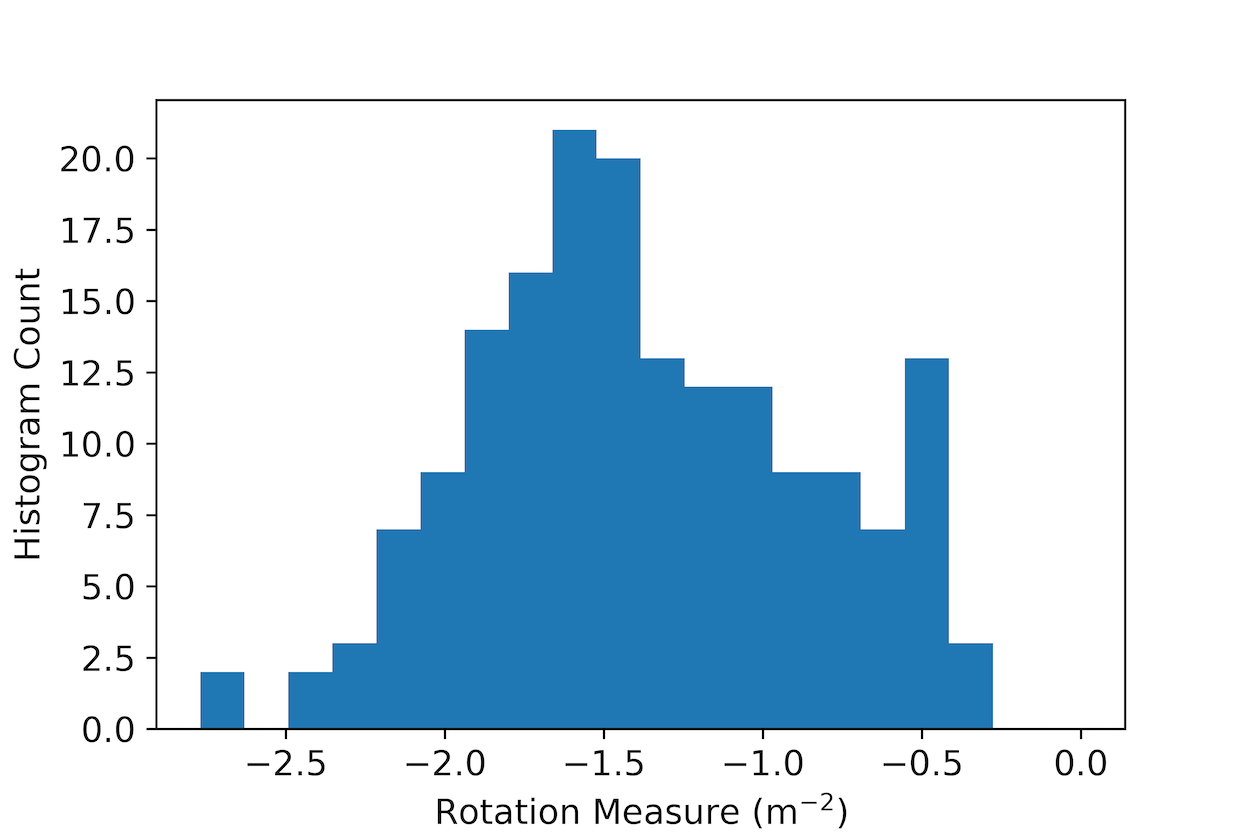}
\label{fig:rm_correction_rm}
}
\subfigure[]{
\includegraphics[width=3.3in]{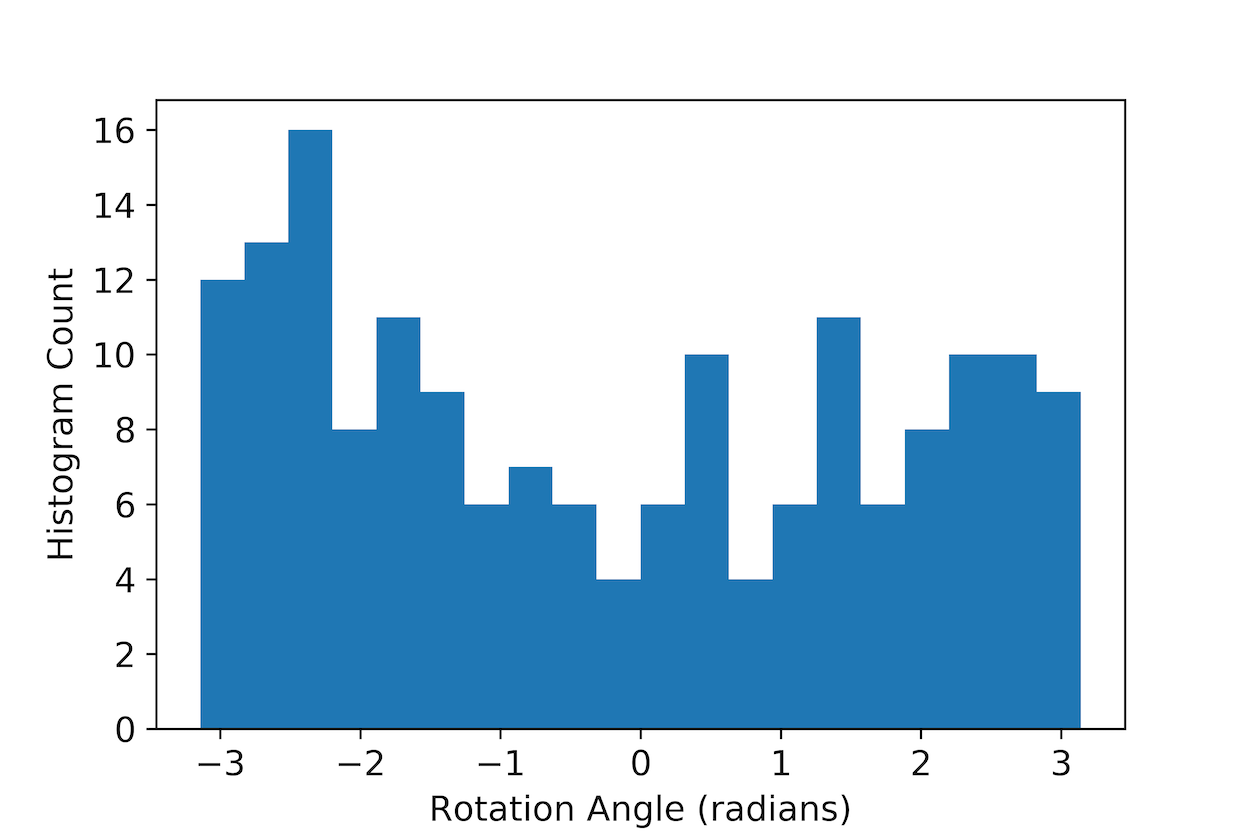}
\label{fig:rm_correction_angle}
}
\caption{Histograms of the applied ionospheric Faraday rotation metrics. The RM is calculated with the \textsc{rmextract} tool. Observations of the ``EoR-0'' field include an empirical correction component, plotted in Figure \ref{fig:rm_empirical}. A single RM is calculated for each of the 172 2-minute observations, plotted in (a). Each RM corresponds to a rotation angle between the Stokes Q and U parameters, as defined by Equation \ref{eq:rm_def} for a single frequency and Equation \ref{eq:eff_rot_angle} for a frequency continuum. (b) histograms the effective rotation angles applied to the data. We find significant Faraday rotation, and all rotation angles are represented in the data.}
\label{fig:rm_correction}
\end{figure}

We assume that ionospheric Total Electron Content (TEC) is constant across an observation's field-of-view. Gradients in the TEC induce positional offsets in compact sources \citep{Loi2015, Jordan2017}, but we can safely ignore these effects when imaging diffuse structure \citep{Lenc2016}. We therefore describe Faraday rotation from the ionosphere with a single RM per observation, plotted in Figure \ref{fig:rm_correction}. The electric field rotation angle $\rotangle$ relates the the RM $\rotmeas$ according to the relationship
\begin{equation}
    \rotangle = \rotmeas \wavelength^2,
\label{eq:rm_def}
\end{equation}
where $\wavelength$ denotes the wavelength.

We estimate the ionospheric RM values with \textsc{rmextract}\footnote{\texttt{https://github.com/lofar-astron/RMextract}}, which calculates the ionospheric TEC along each observation's line-of-sight from Global Position System (GPS) and World Magnetic Model (WMM) data \citep{Mevius2018}. We then correct the continuum Stokes Q and U images for each observation according to the calculation in Appendix \ref{app:rm_calc}. This correction consists of unwrapping the effective rotation angle given by Equation \ref{eq:eff_rot_angle} and boosting the image power by a factor given by Equation \ref{eq:decoherence_factor}, which accounts for depolarization across the full frequency continuum. 

\begin{figure}
\centering
\subfigure[]{
\includegraphics[width=3.3in]{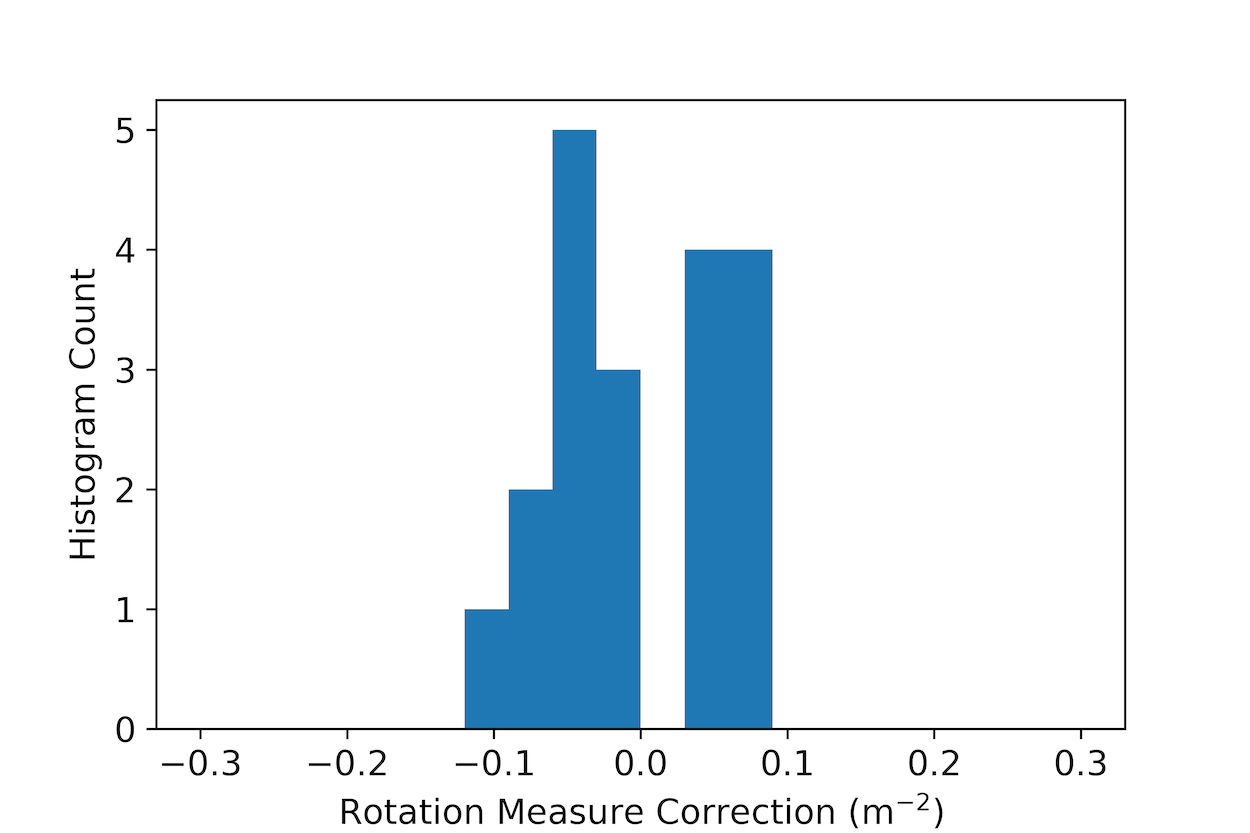}}
\subfigure[]{
\includegraphics[width=3.3in]{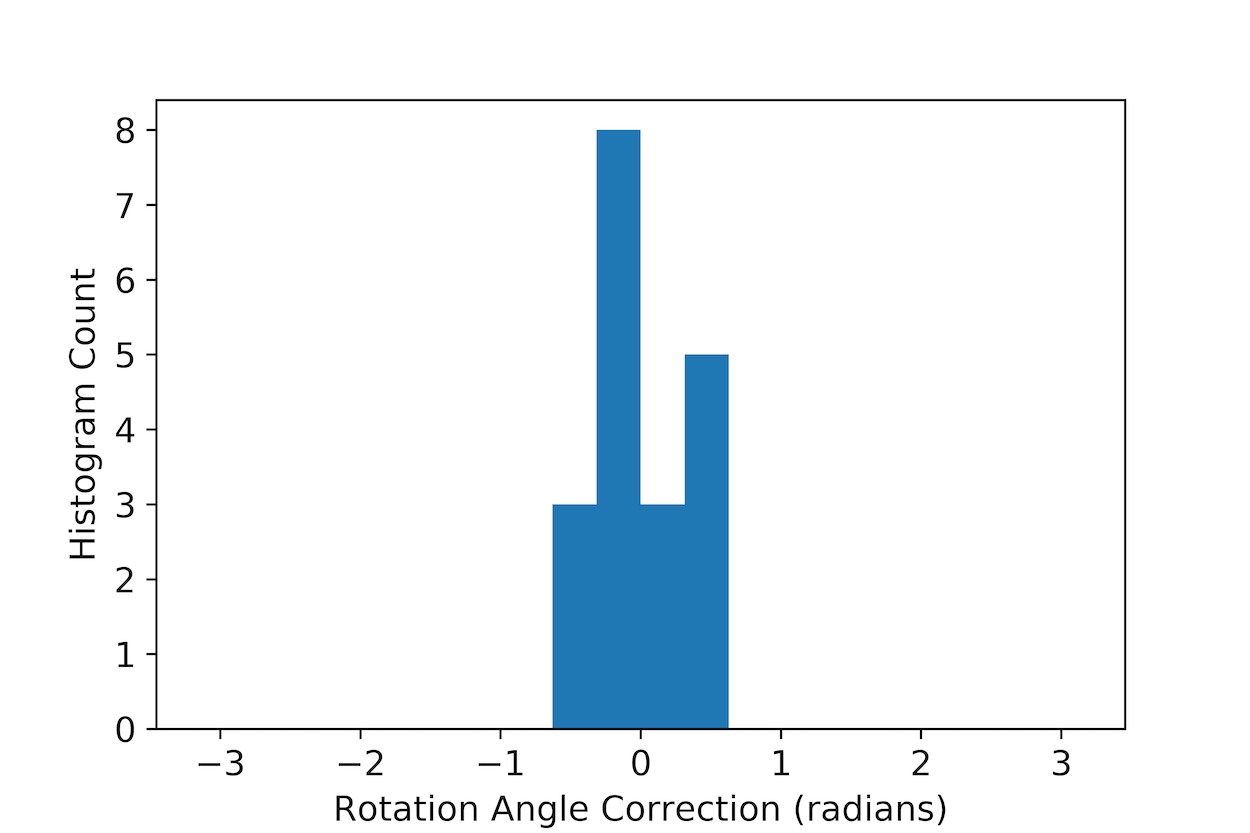}}
\caption{We estimate an empirical self-calibration correction to the RMs of 19 observations near the ``EoR-0'' field. (a) histograms the change in the RM of those observations from this correction. The correction is small compared to the typical RM magnitudes calculated from RMExtract (note that the horizontal axis in (a) has a significantly smaller extent than that of Figure \ref{fig:rm_correction_rm}). (b) presents the associated change in the rotation angles from the RM correction.}
\label{fig:rm_empirical}
\end{figure}

We further correct for Faraday rotation near the EoR-0 field centred at RA 0 h, Dec.\ $-27^\circ$ using an ionospheric self-calibration approach based on that described in \citealt{Lenc2017}. For each observation $\obsindex$, we calculate an empirical correction $\Delta \rotangle_\obsindex$ to the effective rotation angle between Stokes Q and U that maximizes the agreement of that observation to the average map. We minimize the cost function between the average Stokes Q and U maps, $Q_\text{avg}(\skypos)$ and $U_\text{avg}(\skypos)$, and the individual observation's Stokes Q and U maps, $Q_\obsindex(\skypos)$ and $U_\obsindex(\skypos)$:
\begin{equation}
\begin{split}
    \chi^2(\Delta \rotangle_\obsindex) = \sum_{\skypos} & \left( \left[Q_\text{avg}(\skypos) - Q_\obsindex(\skypos) \cos (2\Delta \rotangle_\obsindex) - U_\obsindex(\skypos) \sin (2\Delta \rotangle_\obsindex)\right]^2 \right. \\
    &\left. + \left[U_\text{avg}(\skypos) + Q_\obsindex(\skypos) \sin (2\Delta \rotangle_\obsindex) - U_\obsindex(\skypos) \cos (2\Delta \rotangle_\obsindex)\right]^2 \right).
\end{split}
\end{equation}
It follows that
\begin{equation}
    \tan (2\Delta \rotangle_\obsindex) = \frac{\sum_{\skypos} \left[ Q_\text{avg}(\skypos) U_\obsindex(\skypos) - U_\text{avg}(\skypos) Q_\obsindex(\skypos) \right]}{\sum_{\skypos} \left[ Q_\text{avg}(\skypos) Q_\obsindex(\skypos) + U_\text{avg}(\skypos) U_\obsindex(\skypos) \right]}.
\end{equation}
After calculating this correction to the effective rotation angle we numerically invert Equation \ref{eq:eff_rot_angle} to calculate the associated RM. This conversion is not one-to-one as many RM values correspond to the same effective rotation angle. We choose the closest solution to the initial RM estimate before self-calibration. The results are plotted in Figure \ref{fig:rm_empirical}.

This self-calibration approach works well in fields with a strong linearly polarized signal. We therefore apply this correction only to 19 observations in the ``EoR-0'' field where we measure particularly strong polarized emission. We do not apply ionospheric self-calibration to the map outside of ``EoR-0''.

After correcting for ionospheric Faraday rotation, we repeat the averaging procedure described in \S\ref{s:normalization_and_averaging} to combine the corrected observations. The results are presented below in \S\ref{s:results} and plotted in Figures \ref{fig:stokesIandV}-\ref{fig:stokes_annotated}.

\section{Results}
\label{s:results}

\begin{sidewaysfigure*}
\centering
\subfigure[Unpolarized (Stokes I) diffuse emission.]{
\includegraphics[width=0.9\columnwidth]{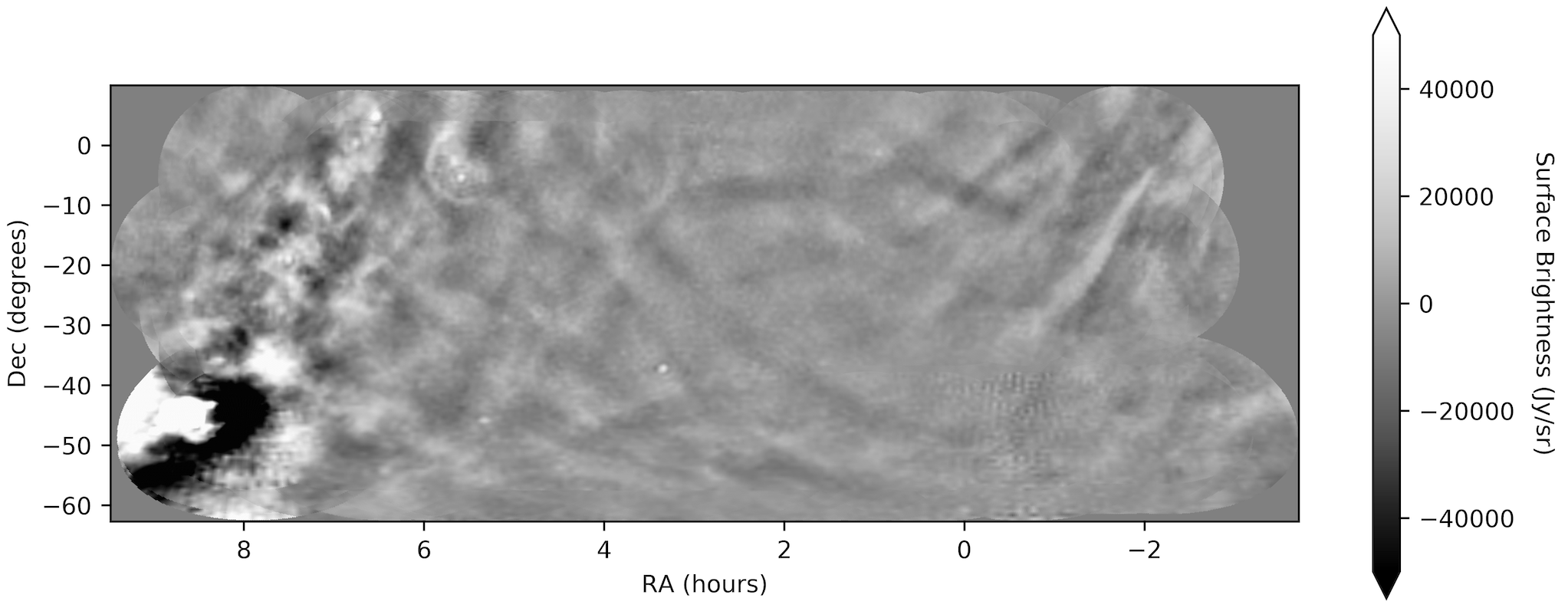}
\label{fig:stokesI}
}
\subfigure[Stokes V circularly polarized diffuse emission.]{
\includegraphics[width=0.9\columnwidth]{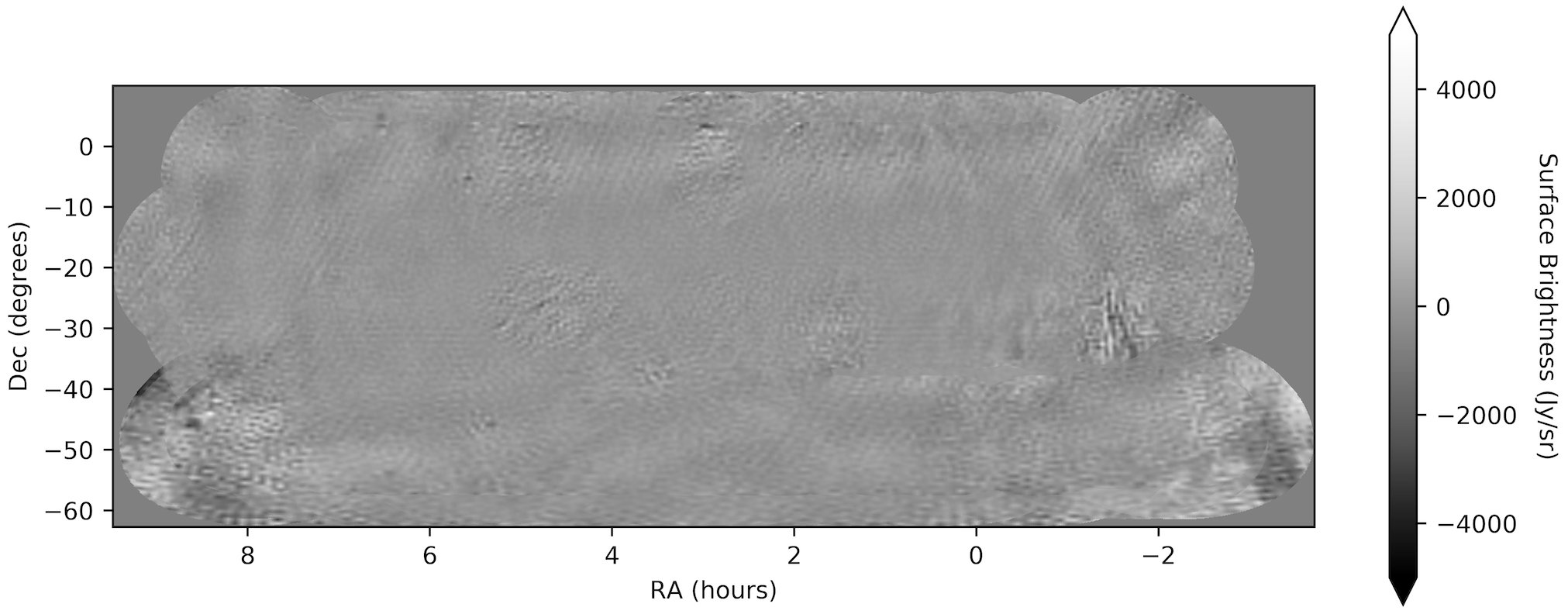}
\label{fig:stokesV}
}
\caption{}
\label{fig:stokesIandV}
\end{sidewaysfigure*}

\begin{sidewaysfigure*}
\centering
\subfigure[Stokes Q linearly polarized diffuse emission.]{
\includegraphics[width=0.9\columnwidth]{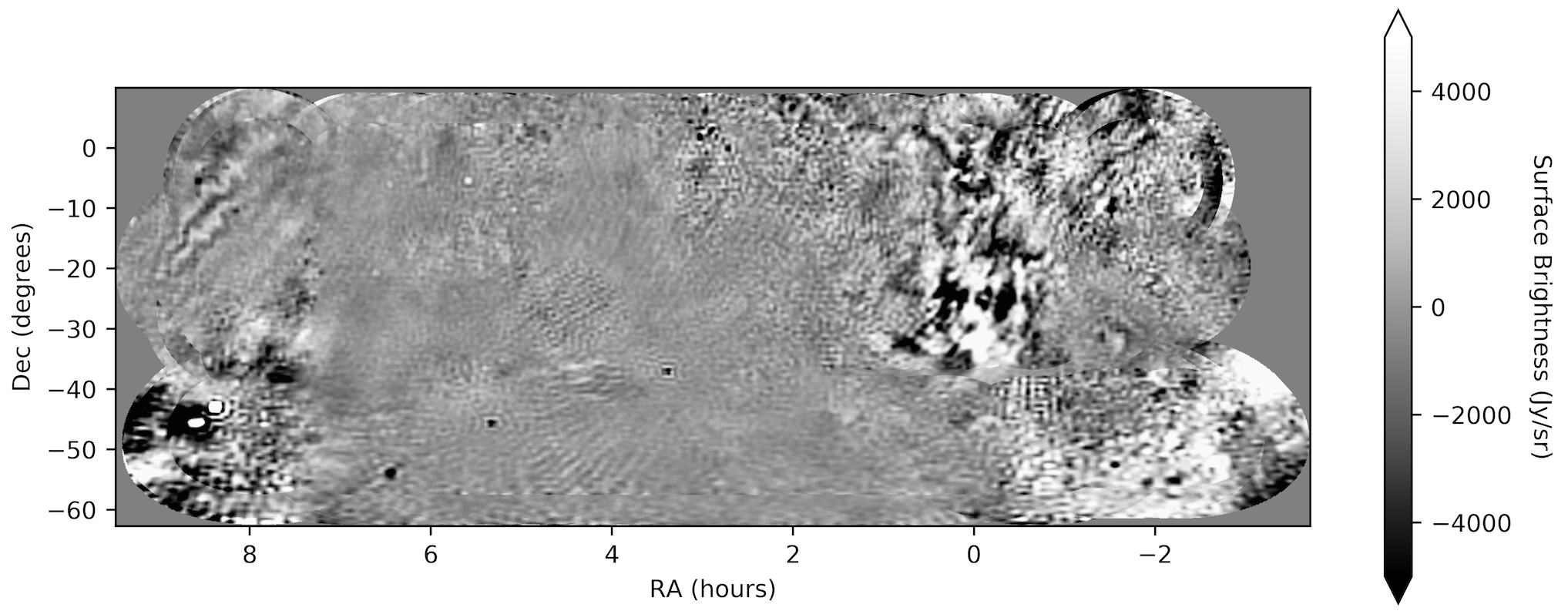}
\label{fig:stokesQ}
}
\subfigure[Stokes U linearly polarized diffuse emission.]{
\includegraphics[width=0.9\columnwidth]{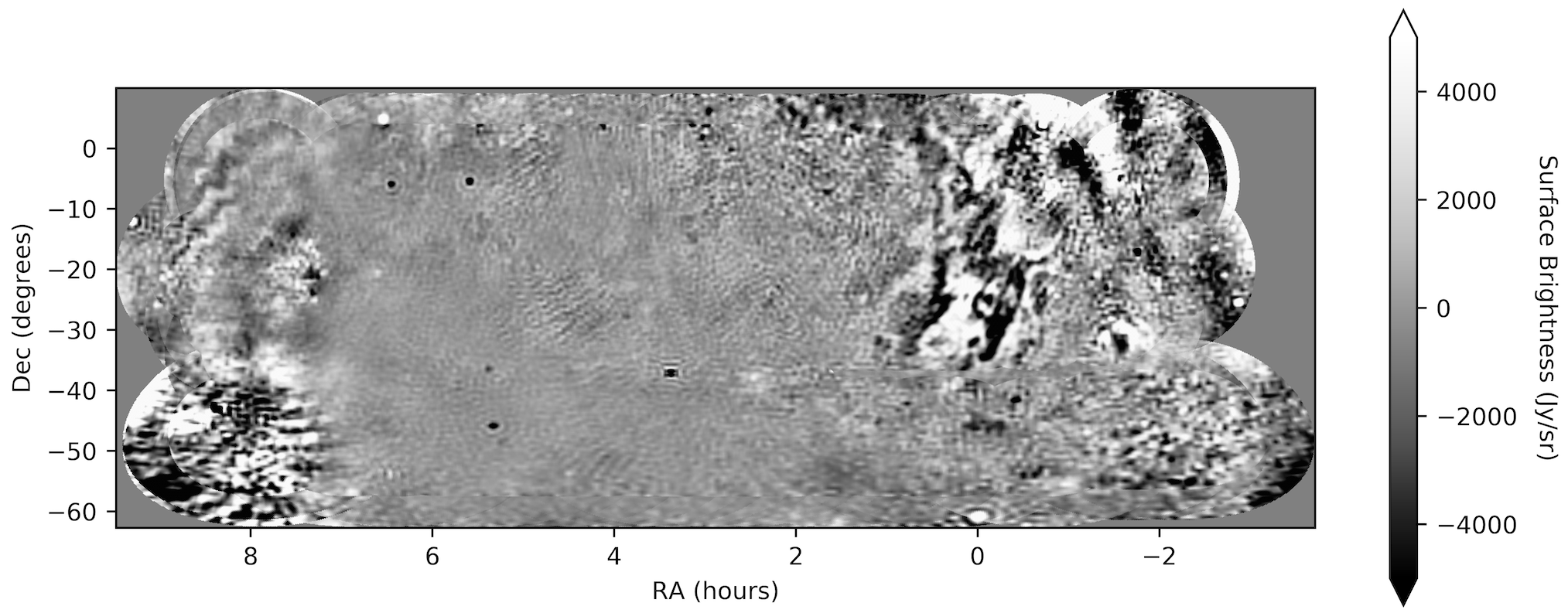}
\label{fig:stokesU}
}
\caption{}
\label{fig:stokesQandU}
\end{sidewaysfigure*}

Figures \ref{fig:stokesIandV} and \ref{fig:stokesQandU} present the polarized mapped diffuse emission. Figure \ref{fig:stokesI} presents the unpolarized (Stokes I) emission, Figure \ref{fig:stokesV} presents the Stokes V circularly polarized emission, and Figures \ref{fig:stokesQ} and \ref{fig:stokesU} present the linearly polarized Stokes Q and U emission, respectively. The Stokes parameters are defined by Equation \ref{eq:stokes_def} above.

The map is presented as a pixelated image with the HEALPix equal-area pixelation scheme \citep{Gorski2005} and is available online as this paper's supplementary material. We provide the map in both \textsc{fits} and \textsc{skyh5} file formats; the \textsc{skyh5} format is compatible with the \textsc{pyradiosky} Python package.\footnote{\texttt{https://github.com/RadioAstronomySoftwareGroup/pyradiosky}}

\begin{figure*}
\centering
\includegraphics[width=1.4\columnwidth]{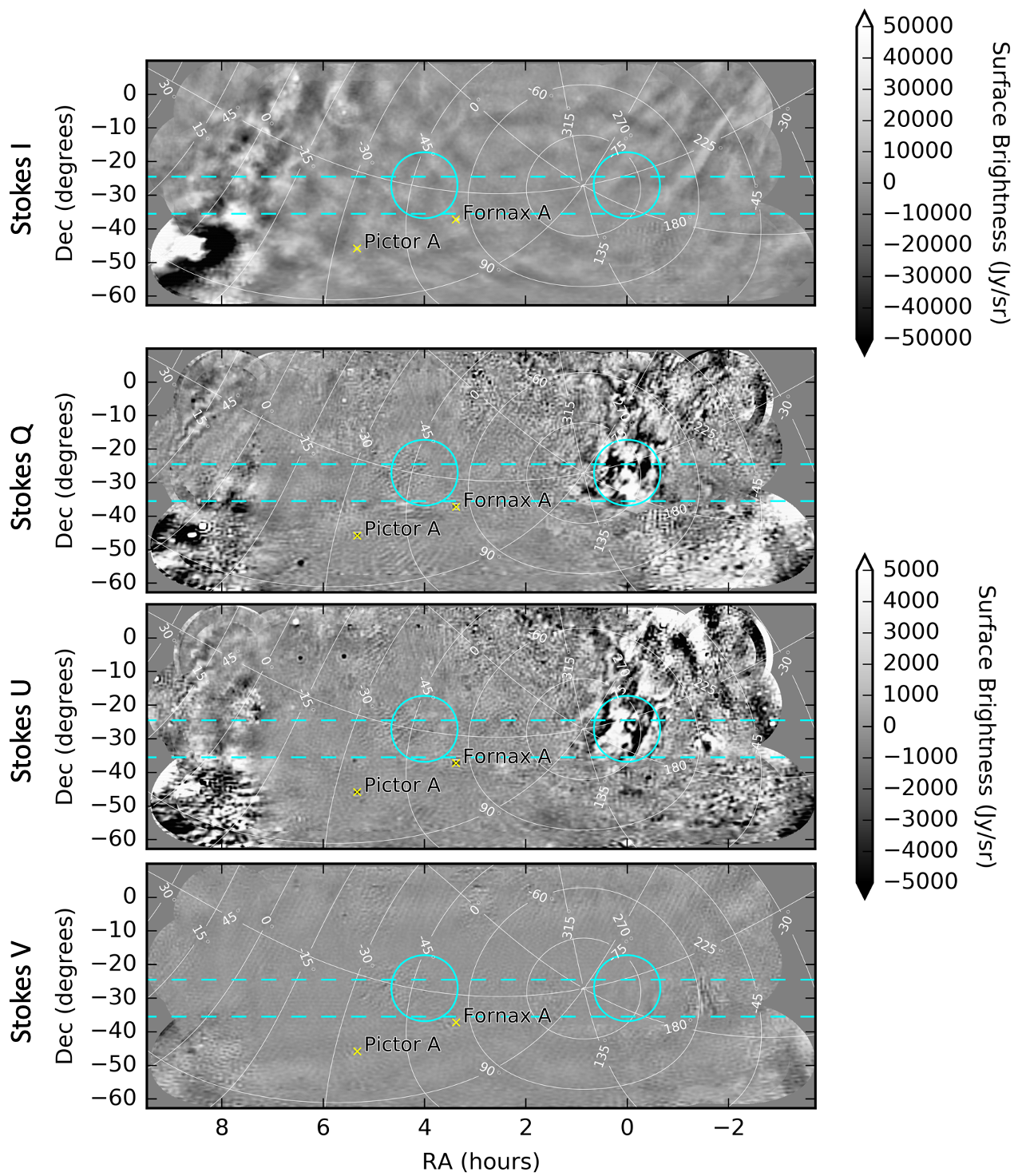}
\caption{Here we reproduce the diffuse map presented in Figures \ref{fig:stokesIandV}-\ref{fig:stokesQandU} with annotations. The dashed cyan lines span the FWHM of the HERA's field-of-view across its observing band, centred at a declination of $-30^\circ$. The solid cyan contours represent the FWHM of the MWA beam at two primary observing fields: ``EoR-0'' is centred at an RA of 0 hours and ``EoR-1'' is centred at an RA of 4 hours. The EoR-0 field contains significant linearly polarized diffuse structure. We label two bright sources, Fornax A and Pictor A, with yellow ``X'' symbols. The map's galactic coordinates are indicated in white.}
\label{fig:stokes_annotated}
\end{figure*}

Figure \ref{fig:stokes_annotated} replicates the map with further annotations. The map covers much of the fields of interest on the sky for EoR science with the MWA and HERA. The solid cyan contours in Figure \ref{fig:stokes_annotated} denote the ``EoR-0'' and ``EoR-1'' fields, regions that have been the focus of the MWA's EoR analyses \citep{Beardsley2016, Barry2019b, Li2019, Trott2020}. The contours are defined as the full width at half maximum (FWHM) of the MWA zenith-pointed beam, averaged across frequencies and polarizations. The dashed cyan lines denote the North and South extents of the FWHM of HERA's beam. Figure \ref{fig:stokes_annotated} also marks two notable bright soures: Fornax A, centred at RA 3.378 h, Dec.\ $-37.21^\circ$; and Pictor A, centred at RA 5.330 h, Dec.\ $-45.78^\circ$. The galactic coordinate system is given in white.

As the map presented here is an interferometric image, we do not measure the absolute amplitude of emission. The map is mean-zero in all Stokes parameters. This presents a number of limitations for our analysis. Although we can confidently measure Stokes polarization across the interferometric modes given by the array design, we cannot calculate the emission's polarization fraction in the image domain. Likewise, we cannot determine the emission's polarization angle.

Even in light of these limitations, we can conclude that the diffuse structure we measure is predominantly unpolarized. Note that the Stokes I image in Figure \ref{fig:stokesI} displays an order of magnitude greater range than the other polarization modes in Figures \ref{fig:stokesV} and \ref{fig:stokesQandU}. We nonetheless measure high levels of linearly polarized diffuse emission. Near RA 0 h, we see very strong polarized signal, consistent with other results in the field \citep{Bernardi2013, Lenc2016}. This structure is particularly concentrated in the MWA's ``EoR-0'' field, suggesting that it could be a significant foreground in EoR analyses such as \citealt{Beardsley2016}, \citealt{Barry2019b}, \citealt{Li2019}, and \citealt{Trott2020}.

We expect the diffuse circularly polarized signal to be very weak and below the sensitivity threshold of this map \citep{Enslin2017, Lenc2018}. Indeed, Stokes V appears largely featureless in Figure \ref{fig:stokesV}. The observed features are imaging artifacts from limitations in calibration precision and polarized beam modeling.

\begin{figure*}
\centering
\includegraphics[width=1.4\columnwidth]{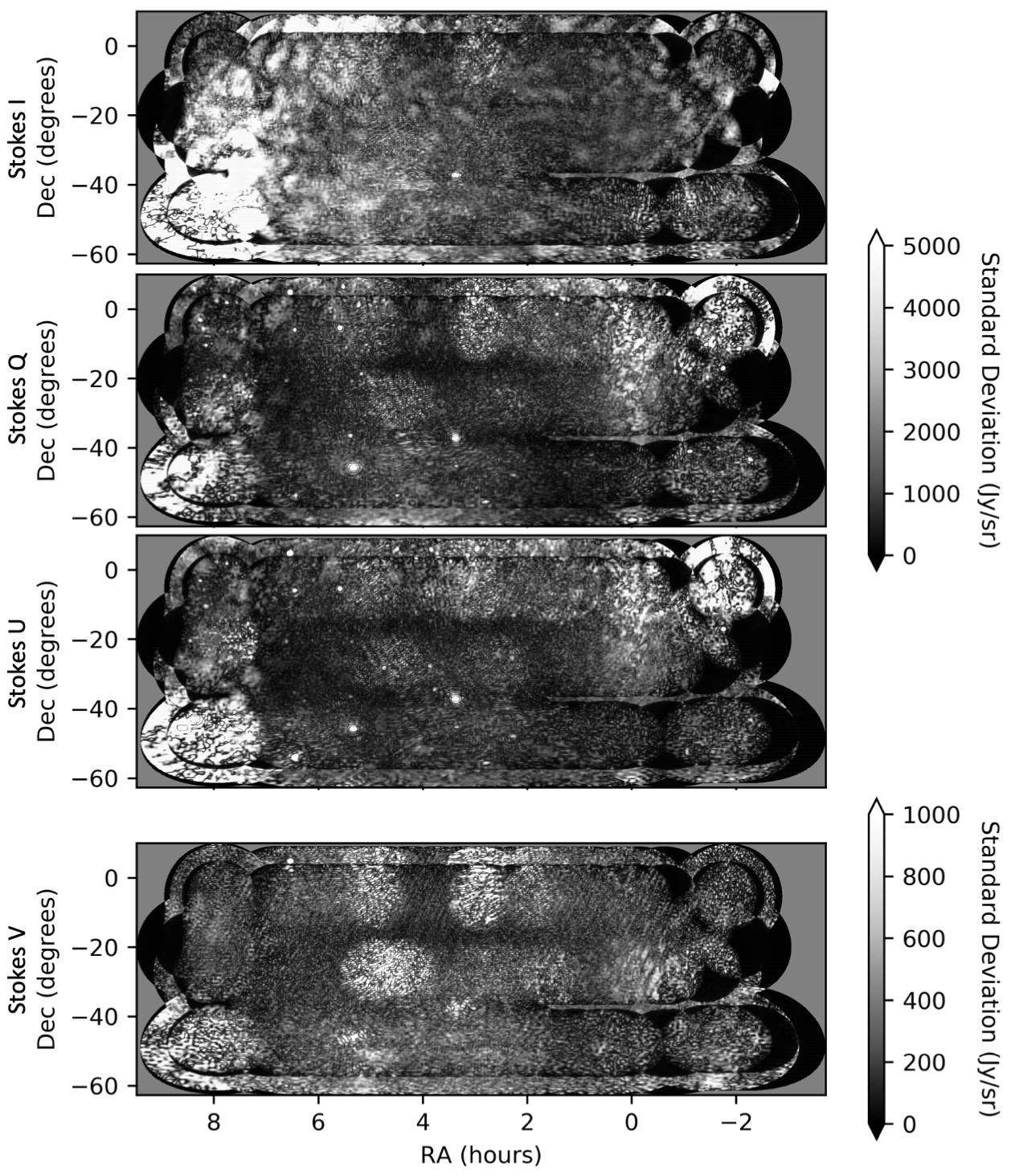}
\caption{Plots of the standard deviation in the mapped diffuse emission across independent observations. For each Stokes parameter and map pixel we calculate the standard deviation across observations that contribute to that pixel measurement via the weighted average described in \S\ref{s:normalization_and_averaging}. High standard deviation values that do not correlate with true diffuse structure indicate systematic errors in the analysis.}
\label{fig:stddev_maps}
\end{figure*}

We measure agreement between overlapping observations by evaluating the standard deviation of each pixel value across contributing observations, plotted in Figure \ref{fig:stddev_maps}. Because the map is mean-zero, we cannot directly compare the standard deviation to the map power to, for example, calculate a signal-to-noise metric. Instead, we present the standard deviation plots in Figure \ref{fig:stddev_maps} alongside the maps in Figures \ref{fig:stokesIandV}-\ref{fig:stokes_annotated} as a qualitative assessment of observation agreement. Features that appear in the standard deviation plots but do not correlate with diffuse structure indicate systematic error. See \S\ref{s:discussion} for further discussion of field-dependent imaging systematics.

\section{Defining Measured Angular Scales}
\label{s:angular_scales}

\begin{figure}
\centering
\subfigure[East-West polarization]{
\label{fig:model_ps_xx}
\includegraphics[height=2.2in]{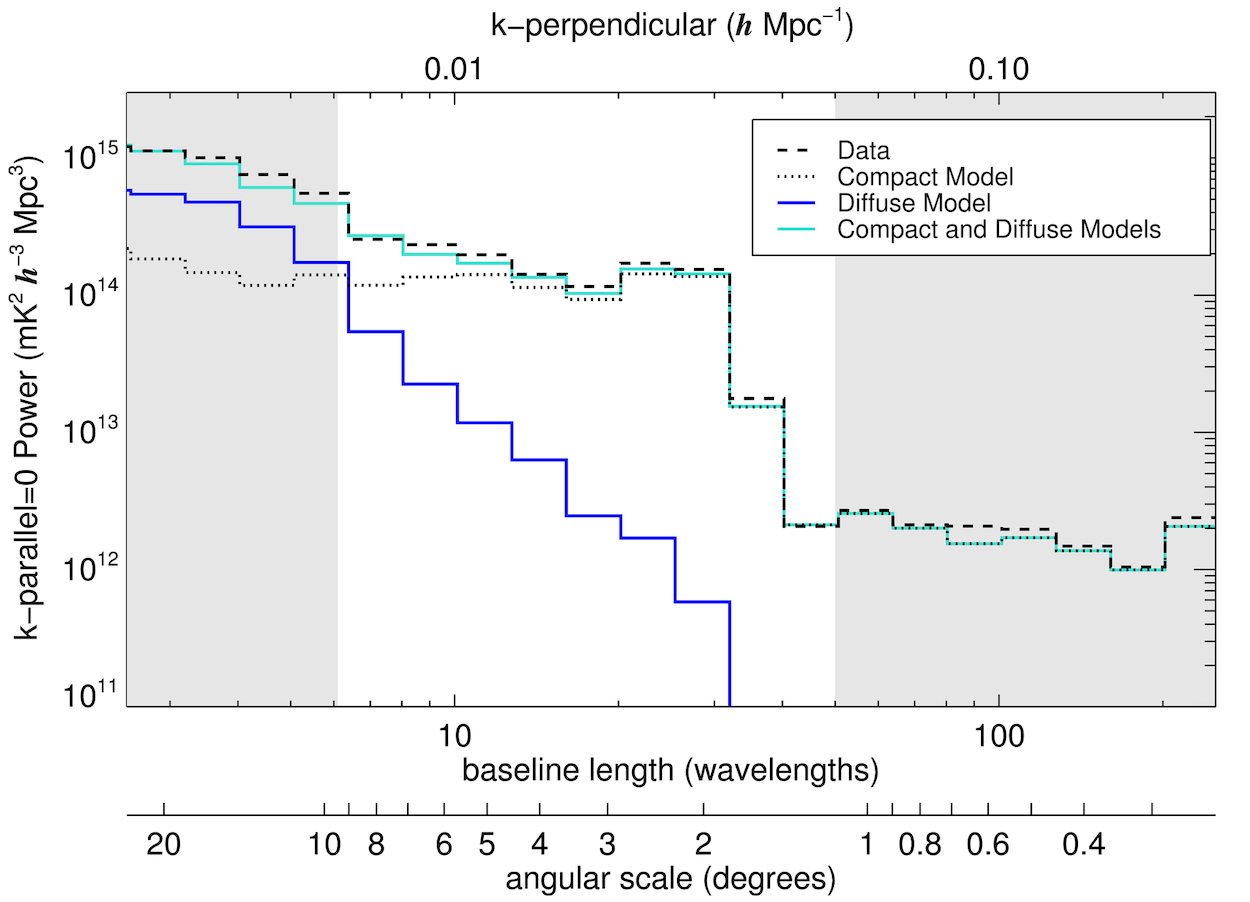}}
\qquad
\subfigure[North-South polarization]{
\label{fig:model_ps_yy}
\includegraphics[height=2.2in]{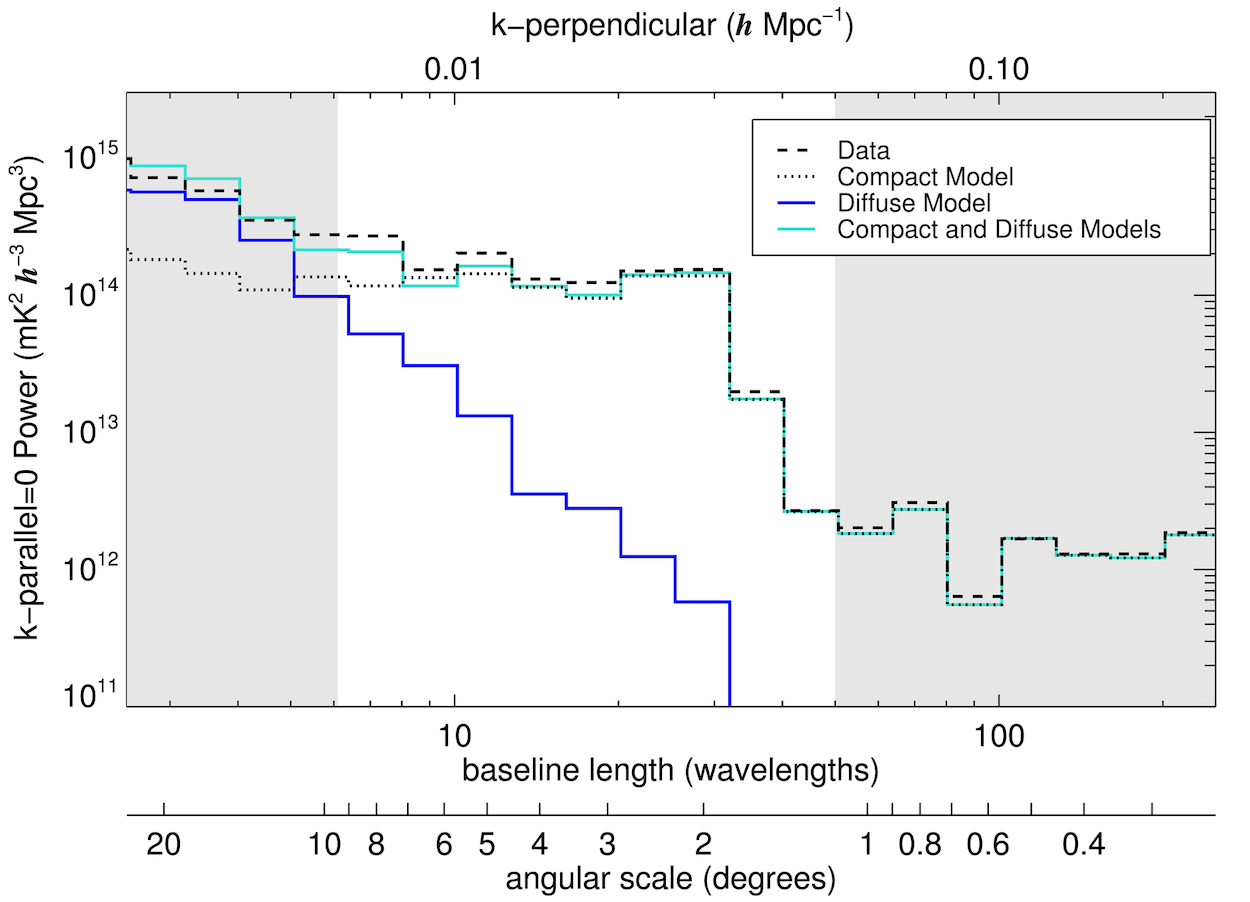}}
\caption{Comparison of the angular power spectra of data and three simulated sky models. The power spectrum represents the frequency-averaged (often denoted $k\text{-parallel}=0$ or $k_\parallel=0$) power at different angular scales on the sky (often denoted $k$-perpendicular or $k_\perp$). The dashed black line corresponds to a single zenith-pointed observation from the MWA of the ``EoR-0'' field. The other three lines correspond to simulations of the same observation based on three different sky models. The dotted black line corresponds to the compact source model derived from the GLEAM catalog \citep{Hurley-Walker2017} and commonly used in calibration (see \S\ref{s:imaging}). The blue line corresponds to the polarized diffuse model presented in \S\ref{s:results}, and the cyan line corresponds to the combined model derived by adding the compact model to the diffuse map. (a) corresponds to the East-West aligned dipoles, and (b) corresponds to the North-South aligned dipoles. We find that the diffuse map recovers power at large scales on the sky that is missing from the compact source model. The shaded gray regions represent scales on which we do not have confidence in the diffuse map. On the right side of each plot, angular scales measured by baselines longer than 50 wavelengths are omitted from the diffuse map presented in this paper. On the left, measurements shorter than 6.1 wavelengths are expected to exhibit bias.}
\label{fig:model_ps}
\end{figure}

Diffuse emission is a significant component of the total sky power at large scales. Figure \ref{fig:model_ps} compares the diffuse map power to that of an MWA observation. The observation, one of the 172 that contribute to this map, is zenith-pointed and centred on the ``EoR-0'' field (centred at RA 0 hours, Dec.\ $-27^\circ$ and plotted in Figure \ref{fig:stokes_annotated}). The dashed black lines present the frequency-averaged angular power spectra of that data. Figure \ref{fig:model_ps_xx} depicts power spectra from the East-West aligned (or $p$) dipoles, and Figure \ref{fig:model_ps_yy} depicts power spectra from North-South aligned (or $q$) dipoles. 

We compare this data to each of three simulations with different sky models. For each sky model, we simulate visibilities for an observation of the ``EoR-0'' field and compute the frequency-averaged angular power spectra. The dotted black lines in Figure \ref{fig:model_ps} correspond to the compact source sky model described in \S\ref{s:calibration} and based on the GLEAM catalog \citep{Hurley-Walker2017}. We see a clear discrepancy between this model and the data at large scales, representing the absence of diffuse structure in the model. The solid blue lines in Figure \ref{fig:model_ps} correspond to the diffuse map presented in \S\ref{s:results}. We include Stokes I, Q, and U emission but omit Stokes V. The solid cyan line represents the sky model derived from combining the diffuse map and the compact source model. We find that the diffuse map recovers the power missing in the compact source model, producing a power spectrum that well-approximates that of the data.

The angular scales captured by the diffuse map are dictated by the MWA's baseline locations. The array's shortest baselines measure large angular scales on the sky, and longer baselines measure small angular scales. Figure \ref{fig:uv_weights} plots the MWA's measurement coverage, with nonzero regions of the plot corresponding to points in the \textit{uv} plane that contribute to one or more visibilities. We see that, below 50 wavelengths (denoted by the large dashed magenta contour in Figure \ref{fig:uv_weights}), nearly every \textit{uv} point is measured. This is a result of the MWA's many closely-spaced antennas and pseudo-random configuration \citep{Beardsley2012}. It also benefits from the large frequency continuum, as baselines measure different \textit{uv} modes at different frequencies.


As noted in \S\ref{s:normalization_and_averaging}, we image only visibility measurements from baselines shorter than 50 wavelengths in order to image in the ``deconvolution-free'' regime. This means that we reconstruct a minimum angular scale of $1.1^\circ$. The map presented in this paper does not include information on angular scales smaller than $1.1^\circ$.

The largest angular scales in the map are limited by the MWA's shortest baselines. The MWA's shortest baseline is centred at 7.7 m in length, equal to 4.7 wavelengths at 182 MHz. However, because of the size of the baseline response kernel, the MWA samples \textit{uv} modes below the centre of this shortest baseline. From Figure \ref{fig:uv_weights}, we find that short baselines sample the \textit{uv} plane down to $\sim 2$ wavelengths. That said, all visibilities that sample the \textit{uv} plane at 2 wavelengths derive from longer baselines. A well-measured \textit{uv} location is measured by an assortment of baselines distributed around the point itself. We therefore expect that measurements of the \textit{uv} plane at 2 wavelengths experience bias.

Because of this sampling bias near the origin of the \textit{uv} plane, we make a conservative estimate of the \textit{uv} modes that are well-measured. We use the threshold down to which all \textit{uv} locations are sampled (2.2 wavelengths) and add the radius of the baseline response kernel (3.9 wavelengths at 182 MHz) to derive an expectation that the \textit{uv} plane is well-measured above 6.1 wavelengths. This corresponds to angular scales up to $9.4^\circ$. However, as we do not filter scales above $9.4^\circ$, the map includes some information from larger angular scales that could be of limited utility in certain calibration and simulation applications (see \S\ref{s:discussion} for further discussion of using this diffuse map for calibration and simulation).

We are confident that the map presented in this paper accurately captures angular scales between $1.1^\circ$ and $9.4^\circ$ ($50-6.1$ wavelengths). The map can be used to reconstruct the \textit{uv} plane for visibility simulation on those scales. We note that, when simulating the map's boundaries, the \textit{uv} plane reconstruction can experience bias from aliasing effects. The map may not allow for accurate visibility simulation of observations near its boundaries.


\section{Evaluating Map Accuracy}
\label{s:validation}

\begin{figure}
\centering
\subfigure[East-West polarization]{
\label{fig:frac_ps_xx}
\includegraphics[height=2.2in]{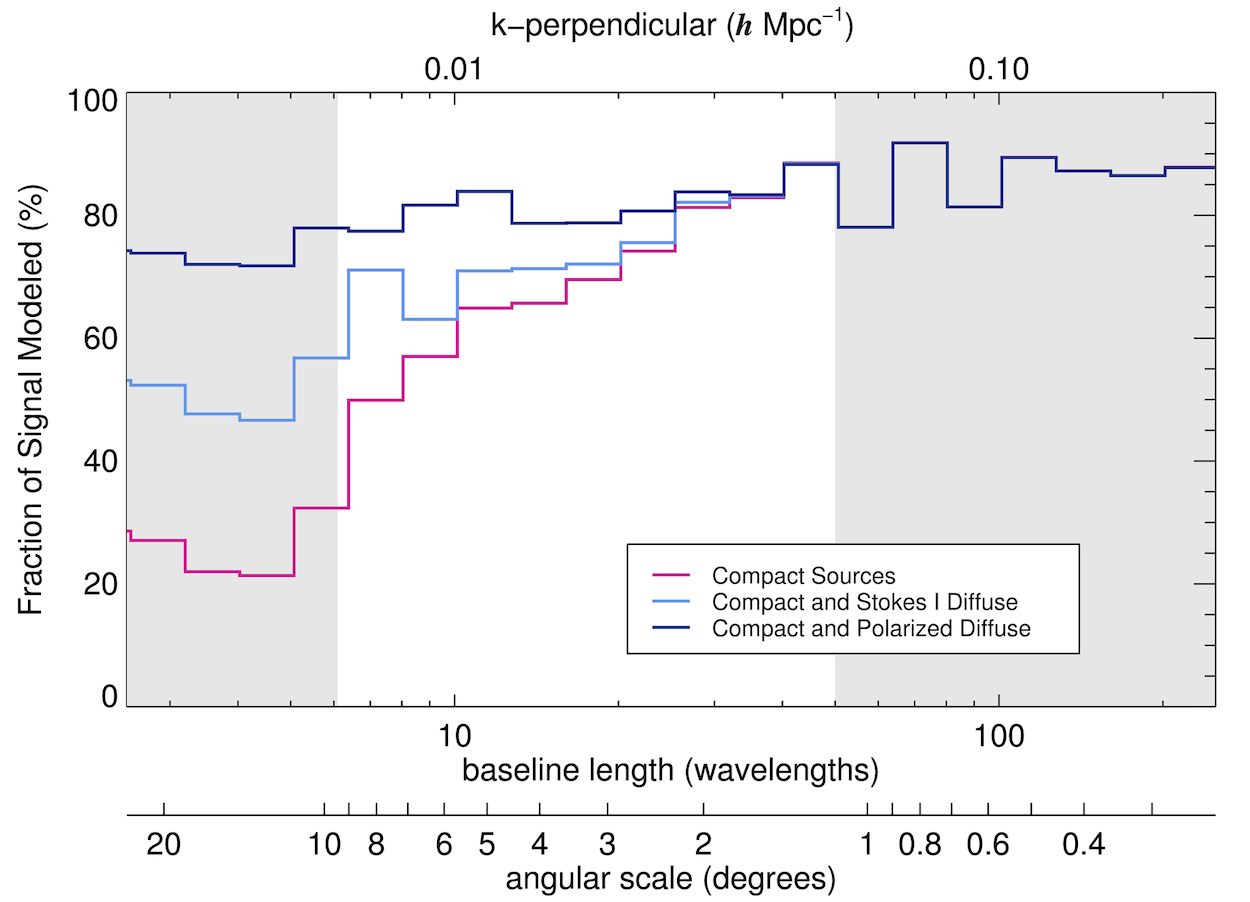}}
\qquad
\subfigure[North-South polarization]{
\label{fig:frac_ps_yy}
\includegraphics[height=2.2in]{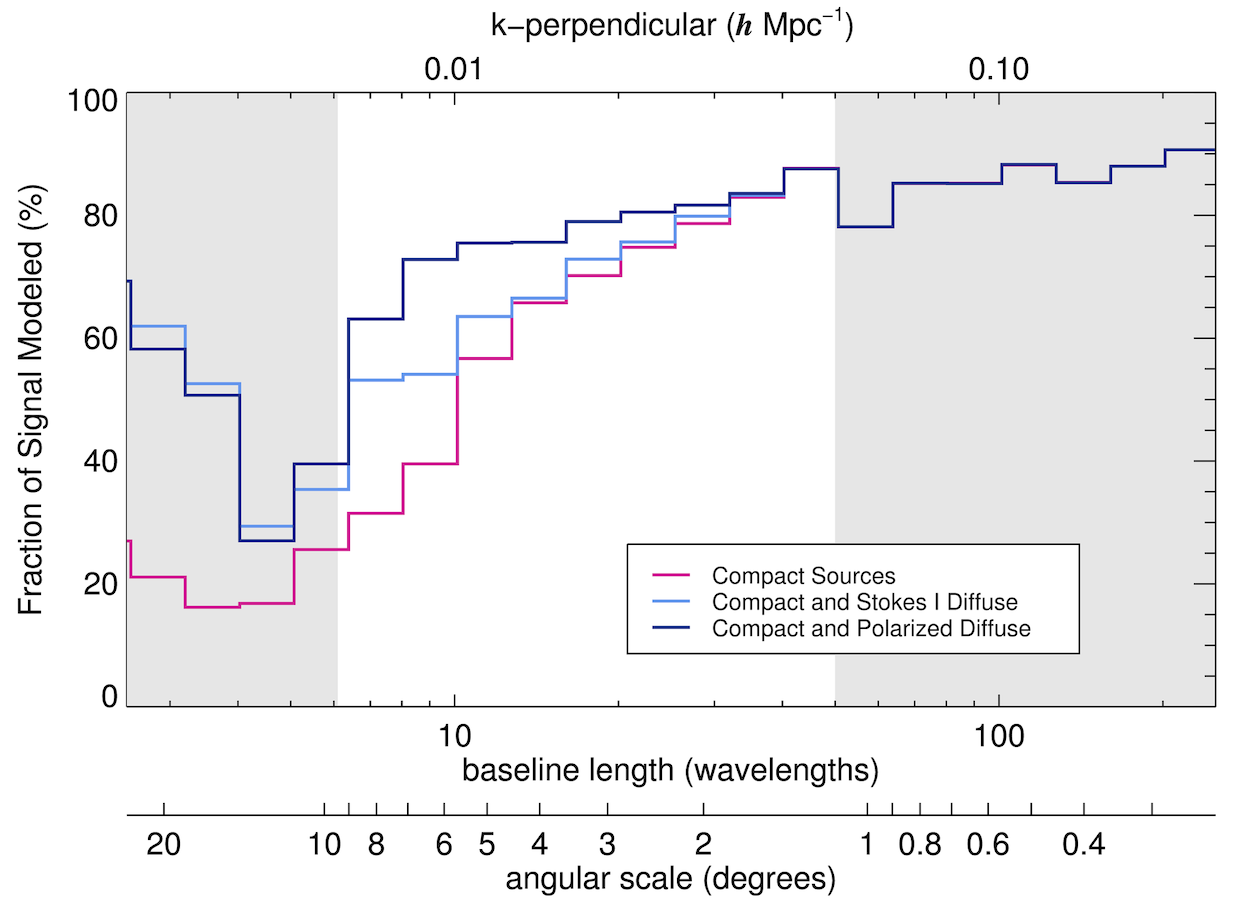}}
\caption{Comparison of the modeling accuracy of one observation with three distinct sky models. The observation is zenith-pointed and centered on the ``EoR-0'' field. The fraction of signal modeled is defined by Equation \ref{eq:fractional_signal_modeled} and quantifies agreement between the measured and modeled visibilities as a function of baseline length. Larger values indicate better modeling accuracy: values of 100\% would indicate that the model perfectly reproduces the visibility measurements for every baseline, time step, and frequency. The magenta lines represent a sky model with compact sources only, the light blue lines include unpolarized diffuse emission in the sky model, and the dark blue lines include polarized diffuse emission. We find that the polarized diffuse map improves modeling accuracy of short baselines to approximately that of long baselines.}
\label{fig:frac_ps}
\end{figure}

The diffuse map is intended to enable accurate visibility modeling for calibration. The most comprehensive validation of the diffuse map's agreement with data is to quantify the discrepancy between measured visibilities and those modeled from the map.

We define a metric of model accuracy as
\begin{equation}
    \text{fraction of signal modeled} = 100\% \times \left(1 - \frac{\operatorname{RMS}\left[ \operatorname{Grid} \left( \data_\text{d} - \data_\text{m} \right) \right]}{\operatorname{RMS}\left[ \operatorname{Grid} \left( \data_\text{d} \right) \right]} \right).
\label{eq:fractional_signal_modeled}
\end{equation}
Here $\data_\text{d}$ are the measured visibilities and $\data_\text{m}$ are the modeled visibilities derived from a sky model. The differenced values $\data_\text{d} - \data_\text{m}$ capture the discrepancy between the data and model and are often called ``residual'' visibilities \citep{Barry2019a, Byrne2019}. ``Grid'' denotes the gridding operation, which transforms the visiblities into a reconstructed \textit{uv} plane, and ``RMS'' denotes evaluating the RMS across \textit{uv} pixels. We calculate the RMS in bands of baseline length to derive a metric of modeling accuracy as a function of baseline length, or angular scale. Values of 100\% would indicate that the modeled visibilities exactly equal the measured visibilities; values of 0\% would indicate very poor agreement, where the discrepancy between the measured and modeled visibilities has equivalent magnitude to the measured visibilities.

In Figure \ref{fig:frac_ps} we use this metric to evaluate the accuracy of modeling a single observation. We use the same zenith-pointed observation of the ``EoR-0'' field plotted in Figure \ref{fig:model_ps}. The plots in Figure \ref{fig:frac_ps} present the results from three sky models: the magenta lines correspond to the compact source model discussed in \S\ref{s:calibration} and based on GLEAM \citep{Hurley-Walker2017}; the light blue lines additionally include unpolarized diffuse emission, plotted in Figure \ref{fig:stokesI}; and the dark blue lines include polarized Stokes Q and U emission as well (Figures \ref{fig:stokesQ} and \ref{fig:stokesU}), which we have corrected to account for the observation's ionospheric rotation measure, as discussed in \S\ref{s:ionosphere}. Figure \ref{fig:frac_ps_xx} presents results from the East-West aligned dipoles, and Figure \ref{fig:frac_ps_yy} presents results from the North-South aligned dipoles. The shaded grey regions represent scales outside our range of confidence for the diffuse map (see \S\ref{s:angular_scales}).

The magenta line in Figure \ref{fig:frac_ps} indicates that, for a sky model consisting of compact sources only, modeling accuracy degrades significantly at short baselines. This metric motivates calibrating with only baselines longer than 50 wavelengths, as discussed in \S\ref{s:calibration}. We find that the diffuse map recovers a significant fraction of the power at large scales, particularly when including polarized emission. For baselines of length 6.1-50 wavelengths, where we are confident in the accuracy of the diffuse map (see \S\ref{s:angular_scales}), the polarized diffuse map improves modeling accuracy to levels commensurate to those of long baselines.

Figure \ref{fig:frac_ps} shows that the diffuse map models a larger fraction of the data in the East-West polarization (Figure \ref{fig:frac_ps_xx}) than in the North-South polarization (Figure \ref{fig:frac_ps_yy}). This effect is most pronounced for baselines shorter than 6.1 wavelengths and may be attributable to the modeling bias discussed in \S\ref{s:angular_scales}. However, we also expect that this effect emerges from the diffuse map's finite extent on the sky. The ``EoR-0'' field is at zenith for this observation, and the bright galactic center is visible above the horizon to the west. The MWA's North-South aligned dipoles are sensitive to low-elevation emission to the east and west and could therefore detect power from the galactic plane that is not captured in our sky model. This points to the diffuse map's limitation in modeling widefield observations near the edges of the mapped region. Future work could measure the galactic plane and other regions beyond the limits of the map presented here.

\section{Discussion}
\label{s:discussion}

The map presented in this paper is intended to enable accurate visibility simulations for a wide range of low-frequency radio instruments, with the particular goal of facilitating sky modeling for precision calibration of 21 cm cosmology measurements. In this section we discuss the process of implementing this map for visibility simulation and calibration.

We have removed compact sources using to the \textsc{fhd} deconvolution algorithm discussed in \S\ref{s:imaging}. To enable accurate visibility modeling, the map must be combined with a source catalog such as the modified GLEAM catalog discussed in \S\ref{s:calibration} or the Long Baseline EoR Survey (LoBES; Lynch et al.\ in review). The success of precision calibration for 21 cm cosmology depends on highly accurate and complete modeling of pointlike and compact extended sources \citep{Carroll2016, Line2020, Zhang2020}.

The diffuse map has physical surface brightness units of Jy/sr, which can be converted to brightness temperature units such as Kelvin. Many visibility simulators represent pixelated images as a series of point sources at the pixel centers. To enable proper normalization, the map must then be converted into units of flux density by multiplying the given values in Jy/sr by the pixel area.

In order to properly model polarized emission in the widefield limit, the visibility simulator must properly convert between the Stokes polarization parameters, defined in Equation \ref{eq:stokes_def}, and the instrumental response using the full direction-dependent instrumental Jones matrix. In the absence of a fully polarized simulator, only the unpolarized Stokes I emission should be used.

As demonstrated in \S\ref{s:validation}, inclusion of Stokes Q and U emission significantly improves visibility modeling in the ``EoR-0'' field, where we observe bright linearly polarized structure. We expect simulations of other fields to benefit less from polarized modeling. In fields with very faint polarized emission, Stokes Q and U could be dominated by imaging errors, in which case visibility simulators could benefit from omitting Stokes Q and U emission entirely. We do not expect modeling Stokes V emission to improve visibility simulations in any field.

The apparent Stokes Q and U emission is highly dependent on ionospheric conditions. If a simulation models this polarized emission, it should account for Faraday rotation from propagation through the ionosphere. At minimum, this should consist of measuring a single ionospheric RM per snapshot observation.

The diffuse map represents continuum observations across a frequency range of 167-198 MHz. It will most accurately represent observations at or near the central frequency of 182 MHz. We have not measured the spectral index or RM of this emission, and it is unclear how to interpolate the map far from this frequency. 

\begin{figure}
    \centering
    \includegraphics[width=3.3in]{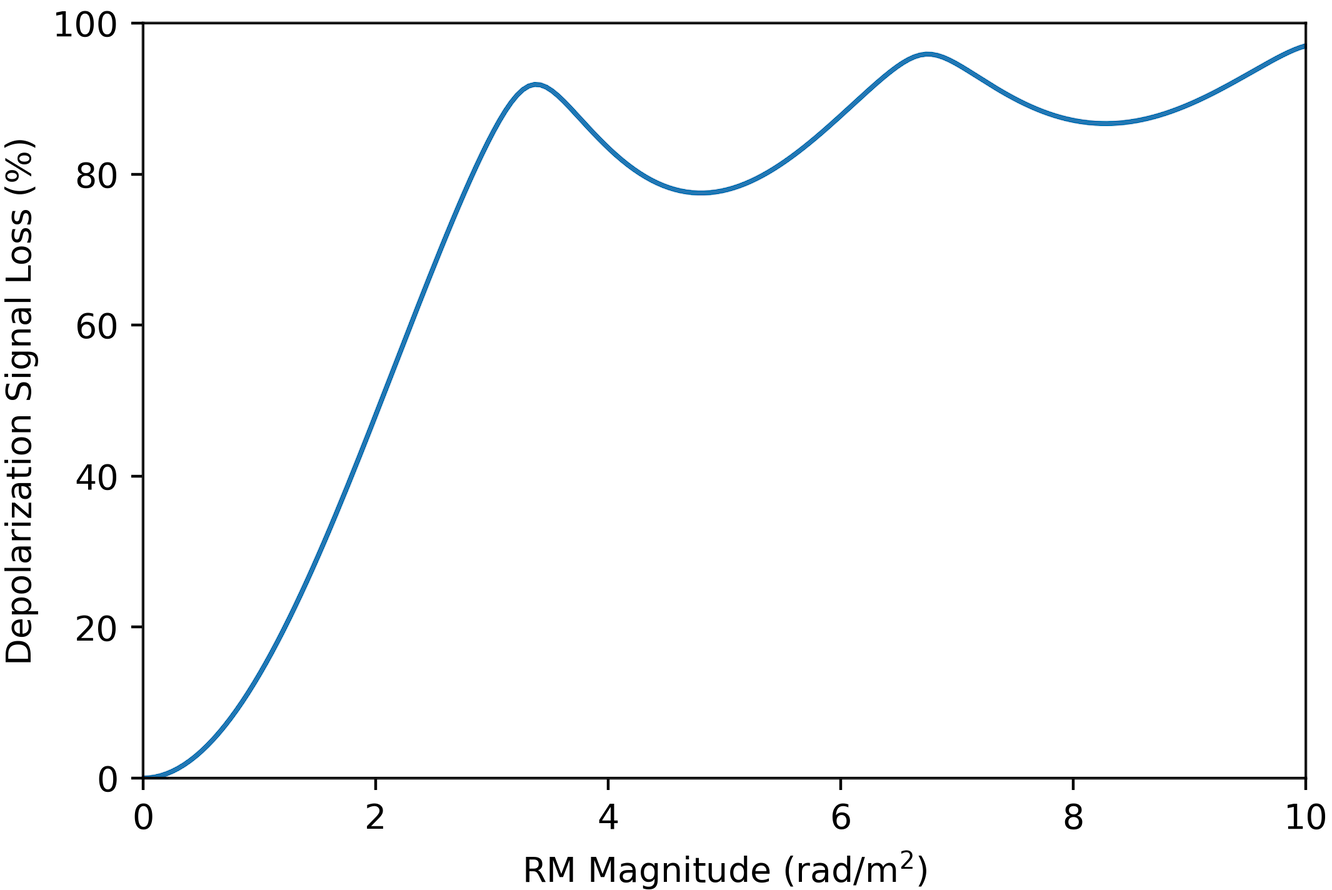}
    \caption{Plot of depolarization as a function of RM magnitude. The diffuse map is measured across a frequency continuum of 167-198 MHz, and in-band depolarization compromises emission with high RM magnitude. Here we plot the fraction of the linearly polarized signal lost, as calculated by Equation \ref{eq:decoherence_factor}. Depolarization increases rapidly with increasing RM magnitude, and by $|R| = 2$ rad/m$^2$ approximately half of the linearly polarized signal is lost.}
    \label{fig:depol}
\end{figure}

Since the map does not capture the diffuse emission's variation across frequency, it cannot accurately model frequency-dependent structure. Emission with high RM magnitude depolarizes across this map's 31 MHz bandwidth, and its polarized features reflect only structure with low RM magnitude. Figure \ref{fig:depol} plots the depolarization fraction as a function of RM magnitude, showing that over half of the linearly polarized signal is lost at RM magnitudes $>2$ rad/m$^2$. There has been significant work in the field exploring the impact of polarized emission with high RM magnitude on the cosmological measurement \citep{Jelic2010, Moore2013, Nunhokee2017, Spinelli2019, Cunnington2021}, but that emission is not captured in this map.

As discussed in \S\ref{s:angular_scales}, the map accurately models angular scales of $1.1^\circ-9.4^\circ$, or \textit{uv} modes between 6.1 and 50 wavelengths. The map does not include information from modes beyond 50 wavelengths. It can be used to accurately simulate visibilities from short baselines that only sample \textit{uv} modes above 6.1 wavelengths. Below 6.1 wavelengths, reconstruction of the \textit{uv} plane is susceptible to measurement bias, as discussed in \S\ref{s:angular_scales}. This bias can affect visibility simulations from baselines that sample \textit{uv} modes below 6.1 wavelengths even when the center of the baseline response lies above 6.1 wavelengths. The impact of this bias is highly instrument-specific and could be minimal in certain situations. For example, instruments with large antennas have baselines that sample large swaths of the \textit{uv} plane. Simulations of these arrays will likely be less vulnerable to this sky model measurement bias.

The diffuse map presented here has a finite extent on the sky. Unmodeled emission could introduce errors in the simulated visibilities of widefield observations that exceed the map's boundaries. Even if emission in an instrument's primary beam is well-modeled, modeling error in the far sidelobes can introduce frequency-dependent calibration errors at a level that is significant for EoR science \citep{Barry2016, Pober2016, Byrne2019}. As noted in \S\ref{s:validation}, the bright galactic center, slightly west of the westernmost extent of the map, is of particularly concern for observations of the ``EoR-0'' field. These errors are mitigated by ensuring that any observational fields simulated with this map are well-contained within the map extent.

As mentioned in \S\ref{s:angular_scales}, Fourier aliasing effects from the map boundaries can also contaminate simulated visibilities. For simulation of observations near the map boundaries, this aliasing can introduce errors in the reconstructed \textit{uv} plane.

Imaging artifacts from poorly deconvolved compact sources can produce errors in simulated visibilities. These errors could be mitigated with improved polarized beam modeling and calibration. Of particular concern are imaging errors near Fornax A and Pictor A that appear predominantly in Stokes Q and U. In order to reduce aliasing in the reconstructed \textit{uv} plane, we do not mask those regions. In certain simulation applications, mitigation approaches such as masking, interpolation over the affected regions, and/or omitting polarized emission from the simulation could be beneficial. The appropriate implementation of these techniques depends on the specific instrumental response and simulation procedure.

Other field-dependent systematic errors are illuminated by the standard deviation plots in Figure \ref{fig:stddev_maps}. These plots suggest that some observations perform relatively poorly, producing high standard deviation values. This is likely related to poor calibration and could be improved with better modeling of the widefield polarized beam. We also observe high standard deviation values in regions of low observational coverage (as plotted in Figure \ref{fig:obs_coverage}). This is apparent at the north edge of the Stokes I standard deviation plot and the top right corner of the Stokes Q and U plots. Furthermore, the high standard deviation values in the lower left corner of the Stokes Q and U in Figure \ref{fig:stddev_maps} is suggestive of Stokes I to Q and U polarization leakage in the vicinity of the Vela supernova remnant, also seen in the polarization leakage estimate in Figure \ref{fig:frac_leakage_estimate}. Once again, visibility simulation could benefit from avoiding these fields, although masking runs the risk of introducing aliasing errors.

While the principal aim of this map is to facilitate precision calibration, visibilities simulated from the map can also be subtracted from data to reduce the foreground signal. As the map is frequency-averaged, this will subtract the frequency-invariant component of the foreground signal only. It will not remove foreground emission with high RM magnitude, which depolarizes in this map (see Figure \ref{fig:depol}). Figure \ref{fig:frac_ps} indicates that, in the EoR-0 field, this will nonetheless appreciably improve short baseline foreground subtraction compared to subtraction of compact sources only.


If implemented properly, we expect this diffuse map to improve calibration precision for cosmological measurements. Results presented in \S\ref{s:validation} indicate that this map enables accurate visibility modeling for baselines as short as 6.1 wavelengths in length. This significantly expands the baseline range available for calibration. As a result, this diffuse map can allow for better calibration signal-to-noise, enable precision calibration of highly compact arrays, and mitigate the frequency-dependent calibration errors inherent to long baseline calibration \citep{Ewall-Wice2017}. Improved calibration sky models that include this map could facilitate the next generation of EoR power spectrum results, allowing for highly sensitive measurements of the the cosmological 21 cm signal.

\section*{Acknowledgements}

We thank James Aguirre, George Heald, Zachary Martinot, and Michael Wilensky for discussions that directly contributed to this work. This work was directly supported by National Science Foundation grants AST-1613855, 1506024, 1643011, and 1835421. NB and CL are supported by the Australian Research Council Centre of Excellence for All Sky Astrophysics in 3 Dimensions (ASTRO 3D), through project number CE170100013. This scientific work makes use of the Murchison Radio-astronomy Observatory, operated by the Commonwealth Scientific and Industrial Research Organisation (CSIRO). We acknowledge the Wajarri Yamatji people as the traditional owners of the Observatory site. Support for the operation of the MWA is provided by the Australian Government (NCRIS), under a contract to Curtin University administered by Astronomy Australia Limited. We acknowledge the Pawsey Supercomputing Centre which is supported by the Western Australian and Australian Governments.

\section*{Data Availability}

The data products underlying this article are available in its online supplementary material. For ease of access, we provide the polarized diffuse maps in two file formats: a \textsc{fits}
format and a \textsc{skyh5} format. The \textsc{skyh5} file is compatible with \textsc{pyradiosky},\footnote{\texttt{https://github.com/RadioAstronomySoftwareGroup/pyradiosky}} a Python utility for reading and writing sky model data and interfacing with simulation packages. The raw data underlying this article were accessed from the MWA All-Sky Virtual Observatory (ASVO)\footnote{\texttt{https://asvo.mwatelescope.org/}} and are publicly available. For instructions on accessing the raw data files from the MWA ASVO, please contact the corresponding author.



\bibliographystyle{mnras}
\bibliography{references} 




\appendix

\section{Faraday Rotation Correction Over a Frequency Continuum}
\label{app:rm_calc}

Given an estimate of the RM $\rotmeas$, we can use Equation \ref{eq:rm_def} to relate observed polarized signal to true emitted signal across a frequency continuum.

At a single frequency $f$, observed Stokes Q and U power $Q'(f)$ and $U'(f)$ relates to true emitted power $Q(f)$ and $U(f)$ via the relationship 
\begin{equation}
    \begin{bmatrix} Q'(f) \\
    U'(f) \end{bmatrix} = \begin{bmatrix}
    \cos[2\rotangle(f)] & -\sin[2\rotangle(f)] \\
    \sin[2\rotangle(f)] & \cos[2\rotangle(f)]
    \end{bmatrix} \begin{bmatrix} Q(f) \\
    U(f) \end{bmatrix}.
\end{equation}
Faraday rotation does not affect Stokes I and V emission.

Continuum images combine signal across a frequency range $(f_\text{min}, f_\text{max})$. The observed continuum Stokes Q and U power $Q'_\text{cont}$ and $U'_\text{cont}$ is given by
\begin{equation}
\begin{split}
    \begin{bmatrix}
    Q'_\text{cont} \\
    U'_\text{cont}
    \end{bmatrix} &=
    \int_{f_\text{min}}^{f_\text{max}} \begin{bmatrix} Q'(f) \\
    U'(f) \end{bmatrix} df \\
    &= \int_{f_\text{min}}^{f_\text{max}} \begin{bmatrix}
    \cos[2\rotangle(f)] & -\sin[2\rotangle(f)] \\
    \sin[2\rotangle(f)] & \cos[2\rotangle(f)]
    \end{bmatrix} \begin{bmatrix} Q(f) \\
    U(f) \end{bmatrix} df.
\label{eq:continuum_rot}
\end{split}
\end{equation}
We now assume that frequency evolution of the observed signal is dominated by Faraday rotation along the propagation path, such that the emitted signal can be approximated as frequency-invariant. Neglecting the frequency dependence of the emitted signal, we approximate the continuum emitted signal as
\begin{equation}
    \begin{bmatrix} Q_\text{cont} \\
    U_\text{cont} \end{bmatrix} = \int_{f_\text{min}}^{f_\text{max}} \begin{bmatrix} Q(f) \\
    U(f) \end{bmatrix} df = (f_\text{max}-f_\text{min}) \begin{bmatrix} Q \\
    U \end{bmatrix},
\end{equation}
where $Q$ and $U$ represent the emitted Stokes Q and U signal at any frequency within the observed frequency range. Plugging this approximation into Equation \ref{eq:continuum_rot}, we get 
\begin{equation}
    \begin{bmatrix}
    Q'_\text{cont} \\
    U'_\text{cont}
    \end{bmatrix} = \frac{1}{(f_\text{max}-f_\text{min})} \left( \int_{f_\text{min}}^{f_\text{max}} \begin{bmatrix}
    \cos[2\rotangle(f)] & -\sin[2\rotangle(f)] \\
    \sin[2\rotangle(f)] & \cos[2\rotangle(f)]
    \end{bmatrix} df \right) \begin{bmatrix} Q_\text{cont} \\
    U_\text{cont} \end{bmatrix}.
\end{equation}

Evaluating, we find that
\begin{equation}
    \begin{bmatrix}
    Q'_\text{cont} \\
    U'_\text{cont}
    \end{bmatrix} = D \begin{bmatrix}
    \cos(2\rotangle_\text{eff}) & -\sin(2\rotangle_\text{eff}) \\
    \sin(2\rotangle_\text{eff}) & \cos(2\rotangle_\text{eff})
    \end{bmatrix} \begin{bmatrix} Q_\text{cont} \\
    U_\text{cont} \end{bmatrix}.
\end{equation}
Here $\rotangle_\text{eff}$ is the effective rotation angle, given by
\begin{equation}
    \rotangle_\text{eff} = \tan^{-1} \left( \frac{\int_{\lambda_\text{min}}^{\lambda_\text{max}} \frac{\sin(2R\lambda^2)}{\lambda^2} d\lambda}{\int_{\lambda_\text{min}}^{\lambda_\text{max}} \frac{\cos(2R\lambda^2)}{\lambda^2} d\lambda} \right).
\label{eq:eff_rot_angle}
\end{equation}
$D$ is a factor that accounts for the decoherence of the signal across the frequency range. It is equal to
\begin{equation}
    D = \frac{1}{\left(\frac{1}{\lambda_\text{min}} - \frac{1}{\lambda_\text{max}}\right)} \sqrt{\left(\int_{\lambda_\text{min}}^{\lambda_\text{max}}\frac{\cos(2R\lambda^2)}{\lambda^2} d\lambda \right)^2 + \left(\int_{\lambda_\text{min}}^{\lambda_\text{max}}\frac{\sin(2R\lambda^2)}{\lambda^2} d\lambda \right)^2}.
\label{eq:decoherence_factor}
\end{equation}
The integrals evaluate to
\begin{equation}
\begin{split}
    \int_{\lambda_\text{min}}^{\lambda_\text{max}} & \frac{\sin(2R\lambda^2)}{\lambda^2} d\lambda = \frac{\sin(2R \lambda_\text{min}^2)}{\lambda_\text{min}} - \frac{\sin(2R \lambda_\text{max}^2)}{\lambda_\text{max}} \\
    &- 2 \sqrt{\pi R} \left[ C(2 \sqrt{R/\pi} \lambda_\text{min}) - C(2 \sqrt{R/\pi} \lambda_\text{max}) \right]
\end{split}
\end{equation}
and
\begin{equation}
\begin{split}
    \int_{\lambda_\text{min}}^{\lambda_\text{max}} & \frac{\cos(2R\lambda^2)}{\lambda^2} d\lambda = \frac{\cos(2R \lambda_\text{min}^2)}{\lambda_\text{min}} - \frac{\cos(2R \lambda_\text{max}^2)}{\lambda_\text{max}} \\
    &+ 2 \sqrt{\pi R} \left[ S(2 \sqrt{R/\pi} \lambda_\text{min}) - S(2 \sqrt{R/\pi} \lambda_\text{max}) \right],
\end{split}
\end{equation}
where $S$ and $C$ are the Fresnel integral special functions:
\begin{equation}
    S(z) = \int_0^z \cos(\pi t^2/2) dt
\end{equation}
and
\begin{equation}
    C(z) = \int_0^z \sin(\pi t^2/2) dt.
\end{equation}


\bsp	
\label{lastpage}
\end{document}